\documentclass[12pt]{amsart}

\usepackage{epsfig}

\begin{document}

\title[]{\bf Conformal Scalar Propagation on the
Schwarzschild Black-Hole Geometry}

\author[]{George Tsoupros \\
       {\em Private address - Beijing}\\
       {\em People's republic of China}\\
}
\thanks{present e-mail address: landaughost@hotmail.com}

\begin{abstract}

The vacuum activity generated by the curvature of the Schwarzschild
black-hole geometry close to the event horizon is studied for the
case of a massless, conformal scalar field. The associated
approximation to the unknown, exact propagator in the Hartle-Hawking
vacuum state for small values of the radial coordinate above $ r =
2M$ results in an analytic expression which manifestly features its
dependence on the background space-time geometry. This approximation
to the Hartle-Hawking scalar propagator on the Schwarzschild
black-hole geometry is, for that matter, distinct from all other. It
is shown that the stated approximation is valid for physical
distances which range from the event horizon to values which are
orders of magnitude above the scale within which quantum and
backreaction effects are comparatively pronounced. An expression is
obtained for the renormalised $ <\phi^2(x)>$ in the Hartle-Hawking
vacuum state which reproduces the established results on the event
horizon and in that segment of the exterior geometry within which
the approximation is valid. In contrast to previous results the
stated expression has the superior feature of being entirely
analytic. The effect of the manifold's causal structure to scalar
propagation is also studied.

\end{abstract}

\maketitle


{\bf I. Introduction}\\

Any reference to the absence of a preferred physical vacuum state in
curved space-times is a statement of fact. An immediate consequence
of that fact is that on curved manifolds the Feynman propagator -
which in Minkowski space-time relates centrally to the dynamical
behaviour of the associated quantum field - ceases to have central
significance. Instead, the quantity which primarily reveals
information about the dynamical behaviour of quantum fields in
curved space-times as well as about the evolution of the space-time
geometry is the vacuum expectation value of the stress-energy tensor
$ <T_{\mu\nu}(x)>$. Also of importance, in this respect, is the -
closely related - quantity $ <\phi^2(x)>$. These two quantities have
the capacity to contain such information because they are local
constructs. As a general consequence of that fact, although these
two quantities depend on the specific choice of vacuum state, their
divergence structure at the coincidence space-time limit $ x
\rightarrow x'$ is independent of that choice at least in the wide
class of space-times and boundary conditions in which these
quantities admit a singularity structure of the Hadamard form. For
the same reason the short-distance behaviour of the propagator at
the same limit is just as independent of the choice of the vacuum
state \cite{Birrel}. This independence is the underlying reason for
the relation between the unrenormalised $ <T_{\mu\nu}(x)>$ and the
Feynman propagator at the coincidence space-time limit
\cite{Birrel}, \cite{Christensen}, \cite{AnHiSam}. The presence of
event horizons in the space-time geometry lends additional
importance to $ <T_{\mu\nu}(x)>$ and to $ <\phi^2(x)>$, not least,
because of the information which they both contain on the evolution
of such horizons. The presence of event horizons requires that
boundary conditions based on the symmetries and causal structure of
the space-time geometry be imposed for a vacuum state to be
determined. In general, the choice of such boundary conditions is
not unique. For that matter, $ <T_{\mu\nu}(x)>$ and $ <\phi^2(x)>$
are themselves defined by the vacuum state which is determined by
the choice of such boundary conditions. At once, this choice of
vacuum state also characterises the Feynman propagator in that
space-time geometry.

It is primarily because of its relation to the stated local
expressions that the Feynman propagator remains an essential
mathematical construct for the study of vacuum activity on curved
manifolds, especially in black-hole space-times. There are,
nevertheless, intrinsic difficulties associated with the
determination of the propagator on curved manifolds in any vacuum
state. Even in the highly symmetric case of the Schwarzschild
geometry the dependence of the Green functions on the radial
variable is unknown. In turn, several approximation schemes have
been devised the primary objective of which has been the development
of general expressions for the scalar propagator as a prerequisite
for the evaluation of the vacuum expectation value of the
stress-energy tensor. Specifically, Candelas \cite{Candelas} has
used the complete set of normalised basis functions for a massless
scalar field obtained by DeWitt in the static region of the
Schwarzschild black-hole space-time in order to develop general
expressions for the massless scalar propagator corresponding to the
Hartle-Hawking, the Boulware and the Unruh vacuum states
respectively. In these expressions the exact dependence on the
Schwarzschild radial variable remains unknown. An exception is the
expression which he has obtained for the scalar propagator
associated with the Hartle-Hawking vacuum state when one end-point
of propagation is specified on the bifurcation two-sphere of the
event horizon on the maximally extended Schwarzschild manifold. In
that case the exact radial dependence becomes explicit at the price
of the stated severe restriction. In the context of the WKB
approximation Anderson \cite{Anderson}, \cite{AnHiSam} has obtained
a general expression of the scalar propagator in static, spherically
symmetric space-times. With the exception of its asymptotic
behaviour at radial infinity that approximation scheme also features
the same lack of information as to the specific dependence of the
scalar propagator on any region of the Schwarzschild black-hole
geometry. By contrast, Page \cite{Page} has obtained the Gaussian
approximation to the path integral for the thermal, conformal scalar
propagator in static metrics. The expression obtained in that
context is explicit in terms of its dependence on the space-time
variables but represents the propagator only at the semi-classical
limit.

In what follows a distinctive approximation to the conformal scalar
propagator associated with the Hartle-Hawking vacuum state - which
corresponds to a black hole in thermal equilibrium with black-body
radiation - will be developed. The ensuing calculation will
eventuate in the analytic expression

\newpage

$$
D(x_2 - x_1) =
$$

$$
\frac{2}{\beta}\frac{1}{\sqrt{\rho_1\rho_2}}\sum_{l=0}^{\infty}\sum_{m=-l}^{l}Y_{lm}(\theta_2,
\phi_2)Y_{lm}^*(\theta_1, \phi_1)\sum_{p =
0}^{\infty}e^{i\frac{p}{\beta}(\tau_2 -
\tau_1)}\int_{u_0[p]}^{\infty}du\frac{cos[\frac{\pi}{4\beta}(4u + 2p
+ 3)(\rho_2 - \rho_1)]}{\pi^2u^2 + 4(l^2 + l + 1)}
$$

$$
- \frac{2}{\beta^{\frac{3}{2}}}\frac{1}{\sqrt{\rho_1}}\times
$$

$$
\sum_{l = 0} ^{\infty}\sum_{m = -l}^{l}\sum_{p =
0}^{\infty}\int_{u_0'[p]}^{\infty}du\frac{cos[\frac{\pi}{4\beta}(4u
+ 2p + 3)(\beta - \rho_1)]}{\pi^2u^2 + 4(l^2 + l +
1)}\frac{J_p(\frac{2i}{\beta}\sqrt{l^2 + l +
1}\rho_2)}{J_p(2i\sqrt{l^2 + l + 1})}e^{i\frac{p}{\beta}(\tau_2 -
\tau_1)}Y_{lm}(\theta_2, \phi_2)Y_{lm}^*(\theta_1, \phi_1) ~~~;~~
$$

$$
(I.1) \hspace{2in} u_0 >> p ~~~~;~~~ u_0' >> p ~~~;~~~ \frac{\pi
u}{\beta}\rho_{2,1} >> p \hspace{3in}
$$
for the conformal scalar propagator in the Euclidean sector of the
Schwarzschild black-hole metric. Unlike all other expressions which
have hitherto been obtained this result constitutes an explicit
expression for the dependence of the propagator on the Schwarzschild
black-hole geometry for values of the radial coordinate $ \rho$ much
smaller than the value $ \rho = 4M$ which labels the non-trivial
boundary of the Riemannian Schwarzschild black-hole manifold. It
will be shown that the analytical extension of (I.1) in real time is
a valid approximation to the Feynman conformal scalar propagator for
values of the Schwarzschild radial coordinate ranging from the event
horizon to an upper bound which is several orders of magnitude above
the range within which particles are spontaneously created and
backreaction effects are pronounced. It will be shown, for that
matter, that there is an ample range of physical distances from the
event horizon within which - in sharp contrast to all previous
results - this Green function, whose space-time dependence is known
and is explicitly featured, is a valid approximation to the exact
conformal scalar propagator.

An essential advantage of the expression in (I.1) is that it renders
the short-distance behaviour and the singularity structure of the
propagator manifest. That expression is, for that matter, especially
suited for an analytic evaluation of $ <T_{\mu\nu}(x)>$ and $
<\phi^2(x)>$. This aspect will be exploited in order to evaluate the
renormalised value of $ <\phi^2(x)>$ in the Hartle-Hawking vacuum
state. In the segment of the Schwarzschild black-hole space-time
within which the approximation for the conformal scalar propagator
is valid the result is

$$
<H|\phi^2(x)|H>_{ren} = \frac{1}{12(8\pi M)^2} \frac{1 -
(\frac{2M}{r})^4}{1 - \frac{2M}{r}} - \frac{1}{32\pi M^2}\frac{1}{(1
- \frac{2M}{r})^{\frac{1}{4}}}\times
$$

$$
\sum_{l = 0} ^{\infty}\sum_{p =
0}^{\infty}\int_{u_0'[p]}^{\infty}du\frac{cos[\frac{\pi}{4}(4u + 2p
+ 3)(1 - \sqrt{1 - \frac{2M}{r}})]}{\pi^2u^2 + 4(l^2 + l +
1)}\frac{I_p[2\sqrt{(l^2 + l + 1)(1 -
\frac{2M}{r})}]}{I_p(2\sqrt{l^2 + l + 1})}(2l + 1)
$$

$$
(I.2) \hspace{2.5in} u_0' >> p \hspace{3in}
$$
and will be shown to be in very good agreement - if not in
coincidence - with the corresponding result obtained in
\cite{CandelHoward} as well as to reduce identically to the
established result on the event horizon obtained in \cite{Candelas}.
However, the result in (I.2) is entirely an analytic function of
space-time. In this respect, within the stated range of validity, it
signifies a better approximation to the renormalised expression of $
<\phi^2(x)>$ than that obtained in \cite{CandelHoward}, since the
latter also involves a numerical component.


In Section II the necessary physical assumptions will be established
and the asymptotic eigenfunctions and eigenvalues to the associated
elliptic operator will be evaluated. In Section III the singular
part which contains the singularity structure of the thermal scalar
propagator will be derived on the basis of the results attained in
Section II. In Section IV the finite, boundary part designed to
cause the propagator to meet appropriate boundary conditions will be
derived. In Section V the validity of the approximation will be
analysed and its range will be established. The singularity
structure of the scalar propagator will be analyzed and the effect
which the causal structure of the Schwarzschild black-hole
space-time has on propagation will be explored. Section VI exploits
the results established in the previous sections in order to
renormalise $ <\phi^2(x)>$ on the black hole's event horizon as well
as in the entire region within which the approximation for the
propagator in (I.1) is valid. The results will be shown to be in
agreement with already established results. Section VII summarises
the main results. As an integral aspect of the calculation
expressions of divergent infinite series as pole structures are
established in the Appendix.

{\bf II. Asymptotic Eigenfunctions and Asymptotic Eigenvalues}\\

The Schwarzschild metric is

\begin{equation}
ds^2 = - (1 - \frac{2M}{r})dt^2 + (1 - \frac{2M}{r})^{- 1}dr^2 +
r^2(d\theta^2 + sin^2\theta d\phi^2)
\end{equation}
The analytical extension $ \tau = + it$ of the real-time coordinate
$ t$ in imaginary values results in a Euclidean, positive definite
metric for $ r > 2M$. The apparent singularity which persists at $ r
= 2M$ can be removed by introducing the new radial coordinate
\cite{SHawking}

\begin{equation}
\rho = 4M(1 - \frac{2M}{r})^{\frac{1}{2}}
\end{equation}
Upon replacing \footnote{The quantity $ 4M$ is denoted by $ \beta$
eventhough it does not correspond to an angle in the Euclidean
Schwarzschild geometry. As only the angle $ 8\pi M = 2\pi\beta$
which corresponds to the temperature of the thermal radiation will
be used in this project this notation is not inconsistent.}

\begin{equation}
\beta = 4M
\end{equation}
the metric in the new coordinates is

\begin{equation}
ds^2 = \rho^2(\frac{1}{\beta^2})d\tau^2 +
(\frac{4r^2}{\beta^2})^2d\rho^2 + r^2(d\theta^2 + sin^2\theta
d\phi^2)
\end{equation}

For the purposes of the calculations herein the Schwarzschild radial
coordinate $ r$ in (4) will be replaced by the Euclidean radial
coordinate $ \rho$ through

\begin{equation}
r = \frac{1}{2}\frac{\beta^3}{\beta^2 - \rho^2}
\end{equation}
obtained by (2). In effect, the metric becomes

\begin{equation}
ds^2 = \frac{1}{\beta^2}\rho^2d\tau^2 + \frac{\beta^8}{(\beta^2 -
\rho^2)^4}d\rho^2 + \frac{\beta^6}{4(\beta^2 - \rho^2)^2}(d\theta^2
+ sin^2\theta d\phi^2)
\end{equation}

The coordinate singularity at $ r=2M$ corresponds to the origin $
\rho = 0$ of polar coordinates and is removed by identifying $
\frac{\tau}{4M}$ with an angular coordinate of period $ 2\pi$.
Equivalently, that coordinate singularity is removed if $ \tau$ is
identified as an angle of period $ 8\pi M$. In addition, although
the curvature singularity at $ r=0$ can not be removed by coordinate
transformations this procedure can be seen to avoid it altogether
along with the entire $ r < 2M$ region of the Schwarzschild geometry
in real time. This procedure results, therefore, in a complete
singularity-free positive definite, Euclidean metric which is
periodic in the imaginary-time coordinate $ \tau$. The period $ 8\pi
M$ of the imaginary-time coordinate $ \tau$ is the underlying cause
of the thermal radiation at temperature $ T = (8\pi M)^{-1}$ emitted
by the Schwarzschild black hole.

The topology of the metric in (6) is $ R^2 \times S^2$. It is
evident from (2) that the value $ \rho = \beta$ is only the label
which this coordinate system assignes to spatial infinity $ r
\rightarrow \infty$. As a consequence of (6) the proper distance
between the origin and $ \rho = \beta$ is infinite. That spatial
infinity accommodates a compact non-trivial boundary of topology $
S^2 \times S^1$ \cite{GibbHawk}.

The Euclidean Green functions for free fields are the unique
solutions of the relevant Green equations which are regular on the
Euclidean section and vanish as $ r \rightarrow \infty$ and,
consequently, as $ \rho \rightarrow \beta$. The Green function of a
massless, conformal scalar field is the solution to

\begin{equation}
[- \square_{x_2} + \xi R(x_2)]D(x_1, x_2) = \delta(x_1, x_2)
\end{equation}
where for the conformally invariant theory it is $ \xi =
\frac{1}{6}$.

Since on the Schwarzschild geometry it is $ R(x) = 0$ this Green
function admits the expansion

\begin{equation}
D(x_2, x_1) = \sum_n\frac{\phi^*_n(x_1)\phi_n(x_2)}{\lambda_n}
\end{equation}
with $ \phi_n$ and $ \lambda_n$ being respectively the
eigenfunctions and eigenvalues of the elliptic operator $ - \square$
on the Euclidean section. That is

\begin{equation}
- \square\phi_n = \lambda_n\phi_n
\end{equation}
The eigenfunctions will be normalised herein by

\begin{equation}
\int d^4x\sqrt{g}\phi^*_n(x)\phi_m(x) = \pi\delta_{nm}
\end{equation}
which differs from the usual orthonormality condition in that the
right side has been scaled by $ \pi$. The reasons for this choice
will be provided in what follows.

An immediate consequence of the fact that the Euclidean
Schwarzschild black-hole geometry has a non-trivial boundary only at
infinite proper distance from the origin is that the spectrum of the
elliptic operator is continuous. For that matter, the summation in
(8) corresponds essentially to an integral and the right side of
(10), to a $ \delta$-function. There are, in addition, no
boundary-related contributions to the propagator for the same
reason. However, in order to keep in line with certain assumptions
made in \cite{SHawking} for numerical purposes the propagator $
D(x_1, x_2)$ will be accorded the discrete representation in (8) and
will be properly replaced by an integral expression at the end of
the calculation on the understanding that no boundary-related
contributions need be calculated.

Separation of variables in (9) yields the eigenfunctions

\begin{equation}
\phi_n = 2\frac{\beta^2 -
\rho^2}{\beta^3}e^{i\frac{p}{\beta}\tau}Y_{lm}(\theta,
\phi)P_{pln}(\rho)
\end{equation}
with the radial function satisfying the equation

\begin{equation}
\frac{1}{\beta^{10}}\rho^2(\beta^2 -
\rho^2)^4\frac{d^2P_{pln}}{d\rho^2} +
\frac{1}{\beta^{10}}\rho(\beta^2 - \rho^2)^3(\beta^2 -
5\rho^2)\frac{dP_{pln}}{d\rho} - V(\rho)P_{pln} =
-\lambda_n\frac{\rho^2}{\beta^2}P_{pln}
\end{equation}
where

\begin{equation}
V(\rho) = (\frac{p}{\beta})^2 + \frac{\rho^2}{\beta^2}[4l(l +
1)(\frac{\beta^2 - \rho^2}{\beta^3})^2  + 4\beta(\frac{\beta^2 -
\rho^2}{\beta^3})^3]
\end{equation}

The parameter $ n$ labels radial eigenvalues. It is discrete if the
Euclidean Schwarzschild black-hole manifold is - by assumption -
bounded and continuous if that manifold is of infinite volume, as is
indeed the case. As the imaginary-time coordinate $ \tau$ has an
angular character the parameter $ p$ is

\begin{equation}
p = 0, \pm1, \pm2, ...
\end{equation}

The expressions given above differ from those in \cite{SHawking} in
that $ r$ has been replaced by $ \rho$ through (5) and, accordingly,
the radial eigenfunction $ R_{pln}(r)$ has been expressed as $
P_{pln}(\rho)$. The resulting

$$
\frac{d^2P_{pln}}{d\rho^2} + \frac{(\beta^2 - 5\rho^2)}{\rho(\beta^2
- \rho^2)}\frac{dP_{pln}}{d\rho} - [4l(l + 1)\frac{\beta^2}{(\beta^2
- \rho^2)^2} + \frac{4}{\beta^2 - \rho^2} +
p^2\frac{\beta^8}{\rho^2(\beta^2 - \rho^2)^4}]P_{pln}
$$

\begin{equation}
= -\lambda_n\frac{\beta^8}{(\beta^2 - \rho^2)^4}P_{pln}
\end{equation}
can be seen to have irregular singular points at $ \rho = \pm\beta$
and, as expected \cite{SHawking}, is quite intractable.

The evaluation of the conformal scalar propagator in the
Hartle-Hawking state commences at this point by developing an
approximate solution to (15). To that effect, propagation on the
Schwarzschild geometry will be examined in the range $ 2M < r < r_0$
which signifies a finite region of the geometry's static segment.
The mathematical consistency of this approximation as well as the
range which $ r_0$ signifies will be analysed and established in
section V and, subsequently, reproduced by comparison to exisiting
results in section VI. Equivalently, on the Euclidean section of the
Schwarzschild metric the evaluation of the scalar propagator will be
made in the context of the assumption


\begin{equation}
\rho^2 << \beta^2
\end{equation}

In effect, (15) reduces to the eigenvalue equation

\begin{equation}
\frac{d^2P_{pln}}{d\rho^2} + \frac{1}{\rho}\frac{dP_{pln}}{d\rho} +
[\lambda_n - \frac{4}{\beta^2}(l^2 + l + 1) -
\frac{p^2}{\rho^2}]P_{pln}(\rho) = 0
\end{equation}
Imposing the condition of regularity at $ \rho = 0$ and expanding in
a power series about that regular singular point yields

$$
P_{pln}(\rho) =
$$

\begin{equation}
(c_0)_{pln}\frac{2^p\Gamma(p + 1)}{\Big{[}\sqrt{\lambda_n -
\frac{4(l^2 + l + 1)}{\beta^2}}\Big{]}^p}
\sum_{k=0}^{\infty}(-)^k\frac{1}{k!4^k2^p} \frac{1}{\Gamma(p + k +
1)}\Big{[}\sqrt{\lambda_n - \frac{4(l^2 + l +
1)}{\beta^2}}\rho\Big{]}^{2k+p}
\end{equation}
with $ p = 0, 1, 2, ...$.

This series converges for all values of $ \rho \in [0, 4M]$ and can
be readily recognised as a Bessel function of the first kind.
Consequently, the solution to (17) is

\begin{equation}
P_{pln}(\rho) = (c_0)_{pln}\frac{2^p\Gamma(p +
1)}{\Big{[}\sqrt{\lambda_n - \frac{4(l^2 + l +
1)}{\beta^2}}\Big{]}^p}J_p\Big{(}\sqrt{\lambda_n - \frac{4(l^2 + l +
1)}{\beta^2}}\rho\Big{)}
\end{equation}

Replacing this expression in

\begin{equation}
\phi_n = \frac{2}{\beta}e^{i\frac{p}{\beta}\tau}Y_{lm}(\theta,
\phi)P_{pln}(\rho)
\end{equation}
- obtained from (11) in the context of (16) - yields

$$
\phi_n(\tau, \rho, \theta, \phi) =
$$

\begin{equation}
(c_0)_{pln}\frac{2^{p + 1}}{\beta}\frac{\Gamma(p +
1)}{\Big{[}\sqrt{\lambda_n - \frac{4(l^2 + l +
1)}{\beta^2}}\Big{]}^p}e^ {i\frac{p}{\beta}\tau}Y_{lm}(\theta, \phi)
J_p\Big{(}\sqrt{\lambda_n - \frac{4(l^2 + l +
1)}{\beta^2}}\rho\Big{)} ~~~;~~~ p = 0, 1, 2, ...
\end{equation}
These are the eigenfunctions to the elliptic operator $ -\square +
\xi R$ on the Euclidean Schwarzschild black-hole geometry in the
context of (16).

It should be remarked at this point that the normalisation condition
\cite{Fawcett}

\begin{equation}
c^2\int_0^{2\pi\beta}|e^{i\frac{p}{\beta}\tau}|^2d\tau = 1
\end{equation}
can also be independently imposed on the temporal sector.
This condition reflects the non-trivial topology of the temporal
sector and does suggest that - at least in the context of (16) -
(10) is more consistent than the usual orthonormality condition.
Indeed, unlike the latter, (10) offsets the effect which the
temporal period has and eventuates in a propagator which features
the familiar power $ \pi^{-2}$ in four dimensions.

It should also be remarked that, as is evident from (17), it is $
[\lambda_n] \sim [\rho^{-2}]$. This is consistent with (9) as well
as with the quantity which appears in the square root in (21) and
shall also test the dimensional consistency of the result which will
be obtained for the eigenvalues.

Since the eigenvalues $ \lambda_n$ appear as a constant in (19)
suitable boundary conditions on some hypersurface will, necessarily,
impose restrictions on their acceptable values. Allowing for the
mathematically trivial condition of regularity imposed throughout
the Euclidean Schwarzschild black-hole geometry the only
hypersurface on which physically meaningful boundary conditions can
be associated with (17) is that at $ \rho = \beta$. Specifically,
since the Euclidean Green functions for free fields are required to
vanish as $ \rho \rightarrow \beta$ the exact eigenvalue equation
(15) which determines the radial sector of the exact eigenvalue
equation (9) must be associated with the boundary condition

\begin{equation}
lim_{\rho \rightarrow \beta}P_{pln}(\rho)= 0
\end{equation}
The eigenvalue equation (17) which is emerges from (15) as a result
of (16) is not, in principle, physically associated with (23).
Mathematically, however, there is no reason that the independent
radial variable $ \rho$ be restricted. Although the solution $
P_{pln}(\rho)$ to (17) is physically relevant only in the context of
(16) the differential equation (17) itself is mathematically
consistent for $ \rho \in [0, \beta]$.
If, for that matter, the boundary condition (23) is also imposed on
the solution to (17) then the latter becomes a good approximation to
the exact radial eigenfunctions which emerge as a solution to (15)
also at $ \rho = \beta$.

In effect, the boundary condition (23) imposed on the eigenfunctions
in (21) is

\begin{equation}
J_p\Big{(}\sqrt{\lambda_n\beta^2 - 4(l^2 + l + 1)}\Big{)} = 0
\end{equation}
This algebraic equation is quite intractable. However, in order to
evaluate the propagator only an asymptotic solution for $ \lambda_n$
is necessary. This is the case because the propagator - being a
Green function - can always be expressed as the sum-total of a
singular part and a smooth function which meets the boundary
conditions.

The evaluation of the asymptotic expression for the eigenvalues
necessitates the asymptotic expression \cite{Arfken}

\begin{equation}
J_{\nu}(z) = \sqrt{\frac{2}{\pi z}}cos[z - (\nu +
\frac{1}{2})\frac{\pi}{2}] ~~~; ~~~ |z| >> |\nu|, -\pi < Argz < \pi
\end{equation}
which - by virtue of the asymptotic condition $ |z| >> |\nu|$ - is,
at once, a necessary and sufficient condition for the eigenfunctions
in (21) to be themselves asymptotic.

As a consequence of (25) equation (24) admits the asymptotic form

\begin{equation}
\sqrt{\frac{2}{\pi}}\frac{1}{\big{[}\lambda_n\beta^2 - 4(l^2 + l +
1)\big{]}^{\frac{1}{4}}}cos\big{[}\sqrt{\lambda_n\beta^2 - 4(l^2 + l
+ 1)} - (p + \frac{1}{2})\frac{\pi}{2}\big{]} = 0
\end{equation}
for values of $ \lambda_n$ high enough to ensure

\begin{equation}
\sqrt{\lambda_n\beta^2 - 4(l^2 + l + 1)} >> p
\end{equation}
so that for these values of $ \lambda_n$ it is

\begin{equation}
\sqrt{\lambda_n\beta^2 - 4(l^2 + l + 1)} - (p +
\frac{1}{2})\frac{\pi}{2} = (n + \frac{1}{2})\pi ~~~; ~~~ n \in Z
\end{equation}
Consequently, the asymptotic solution is

\begin{equation}
\lambda_n = \frac{\pi^2}{16\beta^2}(4n + 2p + 3)^2 +
\frac{4}{\beta^2}(l^2 + l + 1) ~~~; ~~~ n \in Z
\end{equation}
and, as stated, features the expected dimensionality in length
units.

It is evident from (29) that $ \lambda_n$ is asymptotically an
increasing function of $ n$ and that the condition (27) is trivially
satisfied for all values of $ p$ only if

\begin{equation}
n >> p
\end{equation}
Consequently, the asymptotic expression (29) essentially reduces to

\begin{equation}
\lambda_n = \frac{\pi^2}{\beta^2}n^2 + \frac{4}{\beta^2}(l^2 + l +
1) ~~~; ~~~ n \in Z
\end{equation}

This is explicitly an asymptotic expression for the eigenvalues $
\lambda_n$ associated with (17). In fact, as the eigenvalues are
independent of the radial variable $ \rho$ expression (31) also
yields the asymptotic eigenvalues associated with the exact
eigenvalue equation in (15). Consequently, (31) yields the
asymptotic eigenvalues associated with the radial sector of the
elliptic operator $ -\square^2 + \xi R$ on the Euclidean
Schwarzschild black-hole geometry. In this respect it is worth
mentioning that if, in addition to (30), the assumption

\begin{equation}
n >> l
\end{equation}
is made then (31) results in

\begin{equation}
\lambda_n = \frac{\pi^2}{\beta^2}n^2 ~~~; ~~~ n \in Z
\end{equation}
At the limit of infinite radius this expression coincides up to $
\pi^2$ with that obtained in \cite{Fawcett} by placing the black
hole in a spherical box. The condition in (32) reveals that the set
of asymptotic eigenvalues obtained in \cite{Fawcett} is more
restricted than that which corresponds to (31).

In order to arrive at an expression for the asymptotic
eigenfunctions corresponding to the asymptotic eigenvalues in (29)
and (31) the latter must be, respectively, replaced in (21).
Such an operation yields

$$
\phi_n(\tau, \rho, \theta, \phi) = (c_0)_{pln}\frac{2^{p +
\frac{3}{2}}\beta^{p - \frac{1}{2}}}{\pi^{p + 1}}\frac{\Gamma(p +
1)}{\sqrt{n}}\frac{1}{\sqrt{\rho}}e^{i\frac{p}{\beta}\tau}Y_{lm}(\theta,
\phi) \times
$$

\begin{equation}
cos\Big{(}\frac{\pi}{4\beta}(4n + 2p + 3)\rho - p\frac{\pi}{2} -
\frac{\pi}{4}\Big{)} ~~~;~~~ n >> p ~~~;~~~ \frac{\pi n}{\beta}\rho
>> p ~~~;~~~~ p = 0, 1, 2, ...
\end{equation}
on the understanding that the Bessel function in (21) must, itself,
be expressed asymptotically through (25). The condition $ \frac{\pi
n}{\beta}\rho >> p$ is a direct consequence of the asymptotic
condition $ |z| >> |\nu|$ in (25) and will be thoroughly analyzed in
what follows.

In summary, the approximate eigenfunctions of the elliptic operator
$ \square + \xi R$ on the Euclidean black-hole geometry are
expressed in terms of that operator's eigenvalues in (21). This
approximation is valid in the context of (16). The explicit
expression for the asymptotic eigenvalues of the elliptic operator $
\square + \xi R(x)$ is given in (31). The explicit expression for
the asymptotic eigenfunctions of that operator in the context of
(16) is given in (34).

{\bf III. The Scalar Thermal Propagator - Singular Part}\\

In the range of propagation specified on the Euclidean Schwarzschild
black-hole geometry by (16) the evaluation of the conformal scalar
propagator will commence from (8). To that effect, it would appear
that exact knowledge of the eigenvalues $ \lambda_n$ and
eigenfunctions $ \phi_n(x)$ associated with (17) is necessary. This,
however, is not the case. From the outset it is clear that the
replacement of the asymptotic eigenfunctions and eigenvalues
obtained in the preceding section will result in an asymptotic
expression $  D_{as}$ for the propagator. That asymptotic expression
is necessarily the singular part of the Euclidean Green function in
(8). This is the case because the condition in (30) which
characterises the asymptotic eigenfunctions and eigenvalues forces
the lower bound for index n in the infinite series in (8) to
correspond to some $ n_0 >> 1$, an operation which does not affect
the divergence structure which that series has at $ x_2 \to x_1$.
The complete Green function is

\begin{equation}
D(x_2, x_1) = D_{as}(x_2, x_1) + D_{b}(x_2, x_1)
\end{equation}

Apparently, away from the coincidence space-time limit $ x_2
\rightarrow x_1$ the sum-total of the infinite number of terms
corresponding to $ n = 1, n = 2, ... , n = n_0 - 1$ for each
different value of $ p$ respectively yields a finite result. This is
also the case at the coincidence space-time limit since the
singularity in the Green function is necessarily contained in its
asymptotic part $ D_{as}$. Consequently, in the context of (8), the
sum-total of the infinite number of terms corresponding to $ n = 1,
n = 2, ... , n = n_0 - 1$ for each different value of $ p$
respectively corresponds to the boundary part $ D_b$. In effect, $
D_{as}$ satisfies (7) without (necessarily) satisfying the boundary
condition whereas $ D_{b}$ corresponds to an arbitrary smooth
function which satisfies the homogeneous equation associated with
(7) and which can be adjusted to enforce upon $ D(x_2, x_1)$ the
boundary condition at infinity, $ \rho = \beta$.

%
%
%

The evaluation of the asymptotic expression $ D_{as}$ for the scalar
propagator commences with the replacement of (21) into (8) which
yields the expression

$$
D(x_2, x_1) =
$$

$$
\frac{4}{\beta^2}\sum_{l = 0}^{\infty}\sum_{m =
-l}^{l}Y_{lm}(\theta_2, \phi_2)Y^*_{lm}(\theta_1, \phi_1)\sum_{p =
0}^{\infty}e^{i\frac{p}{\beta}(\tau_2 - \tau_1)}\sum_{n =
1}^{\infty}|(c_0)_{pln}|^2\frac{2^{2p}\big{[}\Gamma(p +
1)\big{]}^2}{\Big{[}\sqrt{\lambda_n - \frac{4(l^2 + l +
1)}{\beta^2}}\Big{]}^{2p}}\times
$$

\begin{equation}
\frac{1}{\lambda_n}J_p\Big{(}\sqrt{\lambda_n - \frac{4(l^2 + l +
1)}{\beta^2}}\rho_2\Big{)}J_p\Big{(}\sqrt{\lambda_n - \frac{4(l^2 +
l + 1)}{\beta^2}}\rho_1\Big{)}
\end{equation}
This is the complete propagator associated with (17) expressed in
terms of the eigenvalues $ \lambda_n$ to $ - \square$.

In order to arrive at the explicit asymptotic expression for $
D(x_2, x_1)$ it is necessary to evaluate the coefficients $
(c_0)_{pln}$. Such a task necessitates use of (10). Since the
operator associated with (17) is self-adjoined the orthonormality
relation (10) also applies to the eigenfunctions associated with
(17). Replacing (34) in (10) yields in the asymptotic range given in
(30)

$$
|(c_0)_{pln}|^2\frac{2^{2p + 1}\big{[}\Gamma(p +
1)\big{]}^2}{\big{[}\frac{\pi n}{\beta}\big{]}^{2p}}\times
$$

\begin{equation}
\int_0^{2\pi}d\phi \int_0^{\pi}d\theta
sin\theta\big{|}Y_{lm}(\theta, \phi)\big{|}^2\int_0^{\beta}d\rho\rho
\Big{[}J_p\big{(}\frac{\pi}{4\beta}(4n + 2p + 3)\rho\big{)}\Big{]}^2
= 1 ~~~;~~~ n
>> p
\end{equation}
where use has been made of the invariant-integration measure
associated with the metric in (6).

The double angular integral in (37) is equal to unity on account of
the orthonormality relation

\begin{equation}
\int_0^{2\pi}d\phi\int_0^{\pi}d\theta sin\theta Y_{lm}^{*}(\theta,
\phi)Y_{l'm'}(\theta, \phi) = \delta_{ll'}\delta_{mm'}
\end{equation}
The evaluation of the radial integral in (37) necessitates the
relation \cite{Arfken}

\begin{equation}
\int_0^aJ_{\nu}(a_{\nu m}\frac{\rho}{a})J_{\nu}(a_{\nu
n}\frac{\rho}{a})\rho d\rho = \frac{1}{2}a^2[J_{\nu + 1}(a_{\nu
m})]^2\delta_{mn}
\end{equation}
%
which, as a result of (25), yields

$$
\int_0^{\beta}d\rho\rho\Big{[}J_p\big{(}\frac{\pi}{4\beta}(4n + 2p +
3)\rho\big{)}\Big{]}^2 = $$

\begin{equation}
\frac{\beta^2}{\pi^2}\frac{1}{n} ~~~;~~~ n
>> p
\end{equation}

%

Replacing (40) in (37) yields

\begin{equation}
|(c_0)_{pln}|^2 = \frac{2\pi[\pi n]^{2p + 1}}{[2\beta]^{2p +
2}}\frac{1}{\big{[}\Gamma(p + 1)\big{]}^2} ~~~;~~~ n >> p
\end{equation}

The asymptotic range for the scalar propagator is expected in the
context of the same condition $ n >> p$. The lowest-order term $
n_0$ in the series over $ n$, for that matter, depends on $ p$.
Using (29), (31) and (41) in (36) the asymptotic propagator is

$$
D_{as}(x_2, x_1) = \frac{2\pi^2}{\beta^2}\sum_{l =
0}^{\infty}\sum_{m = -l}^{l}Y_{lm}(\theta_2,
\phi_2)Y^*_{lm}(\theta_1, \phi_1)\times
$$

\begin{equation}
\sum_{p = 0}^{\infty}e^{i\frac{p}{\beta}(\tau_2 - \tau_1)}\sum_{n =
n_0[p]}^{\infty}\frac{n J_p\big{(}\frac{\pi}{4\beta}(4n + 2p +
3)\rho_2\big{)}J_p\big{(}\frac{\pi}{4\beta}(4n + 2p +
3)\rho_1\big{)}}{\pi^2n^2 + 4(l^2 + l + 1)} ~~~;~~~ n_0
>> p
\end{equation}
on the additional condition that - as was the case in (34) - the
Bessel functions in this expression be themselves expressed
asymptotically through (25). Specifically, since - as stated - (25)
is a necessary and sufficient condition for the eigenfunctions in
(21) to be asymptotic it is, at once, a necessary and sufficient
condition for the propagator in (36) to reduce to its asymptotic
expression $ D_{as}(x_2, x_1)$.
To that effect, the condition $ n_0 >> p$ which appears in (42) and
which, as stated, ensures the asymptotic character of the
eigenvalues in (31) must be supplemented with the condition $
\frac{\pi n}{\beta}\rho >> p$. In the context of that condition
expression (42) yields

$$
D_{as}(x_2, x_1) =
\frac{4}{\beta}\frac{1}{\sqrt{\rho_1\rho_2}}\sum_{l =
0}^{\infty}\sum_{m = -l}^{l}Y_{lm}(\theta_2,
\phi_2)Y^*_{lm}(\theta_1, \phi_1)\sum_{p =
0}^{\infty}e^{i\frac{p}{\beta}(\tau_2 - \tau_1)}\times
$$

\begin{equation}
\sum_{n = n_0[p]}^{\infty}\frac{cos\big{(}\frac{\pi}{4\beta}(4n + 2p
+ 3)\rho_2 - p\frac{\pi}{2} -
\frac{\pi}{4}\big{)}cos\big{(}\frac{\pi}{4\beta}(4n + 2p + 3)\rho_1
- p\frac{\pi}{2} - \frac{\pi}{4}\big{)}}{\pi^2n^2 + 4(l^2 + l + 1)}
~~~;~~~ n_0
>> p
\end{equation}
Attention is invited to the fact that, as a result of the condition
$ \frac{\pi n}{\beta}\rho >> p$ the validity of (43) is contingent
upon the radial variable $ \rho$. In fact, the lower bound $ n_0$ in
the series over $ n$ necessarily increases as either $ \rho_2$ or $
\rho_1$ decreases and, indeed, $ n_0$ increases indefinitely as
either radial variable approaches the event horizon at $ \rho = 0$.
The physical repercussions which this situation has to scalar
propagation will be analyzed in due course.

The exponentiation of the cosines in (43) reveals that, at the
coincidence space-time limit at which $ \tau_2 \rightarrow \tau_1 ;
\rho_2 \rightarrow \rho_1$, the expression
%
%
%
%
%

$$
\sum_{p = 0}^{\infty}\sum_{n =
n_0[p]}^{\infty}\frac{e^{i\frac{p}{\beta}(\tau_2 - \tau_1) +
i[\frac{\pi}{4\beta}(4n + 2p + 3)(\rho_2 + \rho_1) - p\pi -
\frac{\pi}{2}]}}{\frac{\pi^2}{\beta^2}n^2 + \frac{4}{\beta^2}(l^2 +
l + 1)} + \sum_{p = 0}^{\infty}\sum_{n =
n_0[p]}^{\infty}\frac{e^{i\frac{p}{\beta}(\tau_2 - \tau_1) +
i[\frac{\pi}{4\beta}(4n + 2p + 3)(\rho_2 -
\rho_1)]}}{\frac{\pi^2}{\beta^2}n^2 + \frac{4}{\beta^2}(l^2 + l +
1)} +
$$

$$
\sum_{p = 0}^{\infty}\sum_{n =
n_0[p]}^{\infty}\frac{e^{i\frac{p}{\beta}(\tau_2 - \tau_1) -
i[\frac{\pi}{4\beta}(4n + 2p + 3)(\rho_2 -
\rho_1)]}}{\frac{\pi^2}{\beta^2}n^2 + \frac{4}{\beta^2}(l^2 + l +
1)} + \sum_{p = 0}^{\infty}\sum_{n =
n_0[p]}^{\infty}\frac{e^{i\frac{p}{\beta}(\tau_2 - \tau_1) -
i[\frac{\pi}{4\beta}(4n + 2p + 3)(\rho_2 + \rho_1) + p\pi +
\frac{\pi}{2}]}}{\frac{\pi^2}{\beta^2}n^2 + \frac{4}{\beta^2}(l^2 +
l + 1)}
$$
diverges. Specifically, the first and the fourth double-series are
convergent whereas the second and the third are divergent. Allowing
for the inconsequential $ l$-dependent constant the second and third
series are, at the stated coincidence limit, of the form

$$
\sum_{p = 0}^{\infty}\sum_{n = n_0[p]}^{\infty}\frac{1}{n^2} =
\frac{1}{n_0^2} + \frac{2}{(n_0 + 1)^2} + \frac{3}{(n_0 + 2)^2} +
... + \frac{n + 1}{(n_0 + n)^2} + ... =
$$

$$
\sum_{n = 1}^{\infty}\frac{n + 1}{(n_0 + n)^2} ~~~;~~~ n_0 >> p
$$
In view of the fact that

$$
\int_1^{\infty}dx\frac{1 + x}{(n_0 + x)^2}
$$
diverges it follows that the second and third double-series also
diverge.

As stated, the parameter $ n$ labels radial eigenvalues and is, for
that matter, continuous because the Euclidean black-hole
Schwarzschild geometry has a non-trivial boundary only at infinite
proper distance from $ \rho = 0$. Consequently, the
logarithmically-divergent integral above replaces identically the
preceding series over $ n$ at the limit at which the boundary's
proper distance from the origin goes to infinity. In effect, that
integration is not merely a formal operation for testing the
convergence properties of the series above but an essential
consequence of the topology of the Euclidean Schwarzschild
black-hole manifold.

The stated first and fourth double-series inherent in (43) can be
absorbed in the finite boundary part $ D_{b}$ of the scalar
propagator. In effect, (43) results in

$$
D_{as}(x_2, x_1) =
$$

$$
\frac{2}{\beta}\frac{1}{\sqrt{\rho_1\rho_2}}\sum_{l =
0}^{\infty}\sum_{m = -l}^{l}Y_{lm}(\theta_2,
\phi_2)Y^*_{lm}(\theta_1, \phi_1)\sum_{p =
0}^{\infty}e^{i\frac{p}{\beta}(\tau_2 -
\tau_1)}\sum_{n=n_0[p]}^{\infty}\frac{cos\big{(}\frac{\pi}{4\beta}(4n
+ 2p + 3)(\rho_2 - \rho_1)\big{)}}{\pi^2n^2 + 4(l^2 + l + 1)}
~~~;~~~
$$

\begin{equation}
n_0 >> p
\end{equation}

%
%
%
%
%

In the context of the stated assumption of a finite proper distance
for the boundary (44) is the asymptotic expression of the massless
conformal scalar propagator. It must be observed that - as a result
of the spherical symmetry of the Schwarzschild manifold - at the
coincidence space-time limit $ x_2 \rightarrow x_1$ the ultraviolet
domain $ n \rightarrow \infty$ enforces the result

$$
D_{as}(x_2 \rightarrow x_1) =
\frac{2}{\pi^2\beta}\frac{1}{\rho}\delta(cos\theta_2 -
cos\theta_1)\delta(\phi_2 - \phi_1)\sum_{p = 0}^{\infty}
\sum_{n=n_0[p]}^{\infty}\frac{1}{n^2} ~~~;~~~
$$

\begin{equation}
n_0 >> p
\end{equation}
Use of

\begin{equation}
\delta(cos\theta_2 - cos\theta_1)\delta(\phi_2 - \phi_1) =
\frac{1}{4\pi}\sum_{k = 0}^{\infty}(2k + 1)P_k(cos\gamma)
\end{equation}
with

$$
cos\gamma = cos\theta_2cos\theta_1 +
sin\theta_2sin\theta_1cos(\phi_2 - \phi_1)
$$
in (45) reveals, by simple power counting, that the singularity
contained in (44) in four dimensions at the coincidence space-time
limit is a quadratic divergence. In real time this divergence will,
of course, also occur when two distinct points $ x_2$ and $ x_1$ can
be joined by a null geodesic. The quadratic divergence inherent in
(44) is consistent with theoretical expectations of a scalar
propagator on any manifold. This aspect of the scalar propagator on
the Schwarzschild black-hole geometry will be further explored in
Section V. It is also worth noting that -  as stated in Sec II - the
singular part of the propagator in (45), consistently, features the
factor $ \pi^{-2}$ which is also expected of a scalar propagator in
four dimensions.

Since the actual spectrum of eigenvalues $ \lambda_n$ is continuous
as a result of the infinite volume of the Euclidean black-hole
geometry the expression for the asymptotic propagator which
corresponds to the actual geometry will be obtained by replacing the
series over $ n$ in (44) by an integral. Consequently,

$$
D_{as}(x_2, x_1) =
$$

$$
\frac{2}{\beta}\frac{1}{\sqrt{\rho_1\rho_2}}\sum_{l =
0}^{\infty}\sum_{m = -l}^{l}Y_{lm}(\theta_2,
\phi_2)Y_{lm}^*(\theta_1,\phi_1)\sum_{p =
0}^{\infty}e^{i\frac{p}{\beta}(\tau_2 -
\tau_1)}\int_{u_0[p]}^{\infty}du\frac{cos[\frac{\pi}{4\beta}(4u + 2p
+ 3)(\rho_2 - \rho_1)]}{\pi^2u^2 + 4(l^2 + l + 1)}~~~;~~~  $$

\begin{equation}
u_0 >> p ~~~ ; ~~~ \frac{\pi u}{\beta}\rho_{2,1} >> p
\end{equation}
where, for notational consistency, the discrete parameter $ n$ has
been replaced by the continuous variable $ u$ the lower limit $ u_0$
of which is also a function of $ \rho_2$ or $ \rho_1$ accordingly.

In the context of (16) this is the asymptotic expression for the
massless conformal scalar propagator in the Hartle-Hawking state.
Equivalently, this is the expression for the singular part of that
propagator.

{\bf IV. The Scalar Thermal Propagator - Boundary Part}\\

As stated, the boundary part $ D_b$ of the propagator satisfies the
homogeneous equation associated with (7)

\begin{equation}
\square_{x_2} D_{b}(x_2 - x_1) = 0
\end{equation}
and can be adjusted to enforce the boundary condition

\begin{equation}
D(x_2, x_1)_{|\rho_2 = \beta \bigvee \rho_1 = \beta} = 0
\end{equation}
of vanishing propagation on the boundary of the Euclidean
Schwarzschild black-hole geometry. This boundary condition is
physically relevant to the propagator $ D$ despite the range of
propagation specified in (16) for the same reasons which lend
validity to (24). Since, as analysed in the previous section, in the
context of (8) the boundary part $ D_b$ is specified by the
sum-total of all terms corresponding to $ n = 1, n = 2, ..., n = n_0
- 1$ for all values of $ l$ and $ p$ it follows that

\begin{equation}
D_{b}(x_2 - x_1) = \sum_{lmp}\sum_{n = 1}^{n_0 -
1}\frac{\phi_{nplm}(x_2)\phi_{nplm}^{*}(x_1)}{\lambda_{nplm}}
~~~;~~~ n_0 >> [p,l]_{max}
\end{equation}
At once, (48) implies that $ D_b$ satisfies (9) with $ \lambda_n =
0$. Consequently, the expression in (50) is independent of $
\lambda_n$. As a result of (48) and (21) the boundary part is

$$
D_{b}(x_2 - x_1) =
$$

\begin{equation}
\frac{1}{\beta}\sum_{p = 0}^{\infty}\sum_{l = 0} ^{\infty}\sum_{m =
-l}^{l}\sum_{n = 1}^{\infty}A_{plmn}\frac{2^{p + 1}\Gamma(p +
1)}{\Big{[}\frac{2i}{\beta}\sqrt{ l^2 + l +
1}\Big{]}^p}e^{i\frac{p}{\beta}\tau_2}Y_{lm}(\theta_2, \phi_2)
J_p(\frac{2i}{\beta}\sqrt{l^2 + l + 1}\rho_2)
\end{equation}

It is worth mentioning that although (50) and (51) are both
expressions for the boundary part $ D_b$ they are very different in
terms of their mathematical essence. Specifically, (50) expresses $
D_b$ as the complement to the singular part $ D_{as}$ of the
propagator whereas (51) expresses $ D_b$ as a linear combination of
those eigenfunctions in (9) which correspond to $ \lambda_n = 0$.

The coefficients $ A_{plmn}$ are expected to manifest an explicit
dependence on the space-time point $ x_1$ and will naturally be
determined by the boundary condition in (49) which, in view of (35),
(44) and (51), becomes

$$
\frac{2}{\beta^{\frac{3}{2}}}\frac{1}{\sqrt{\rho_1}}\sum_{l =
0}^{\infty}\sum_{m = -l}^{l}Y_{lm}(\theta_2,
\phi_2)Y_{lm}^{*}(\theta_1, \phi_1)\sum_{p =
0}^{\infty}e^{i\frac{p}{\beta}(\tau_2 -
\tau_1)}\sum_{n=n_0'[p]}^{\infty}\frac{cos[\frac{\pi}{4\beta}(4n +
2p + 3)(\beta - \rho_1)]}{\pi^2n^2 + 4(l^2 + l + 1)} ~~~~~~ +
$$

\begin{equation}
\sum_{p = 0}^{\infty}\sum_{l = 0} ^{\infty}\sum_{m = -l}^{l}\sum_{n
= 1}^{\infty}A_{plmn}\frac{2\beta^{p - 1}\Gamma(p +
1)}{\Big{[}i\sqrt{l^2 + l + 1}\Big{]}^p}J_p(2i\sqrt{l^2 + l +
1})e^{i\frac{p}{\beta}\tau_2}Y_{lm}(\theta_2, \phi_2) = 0 ~~~;~~~
n'_0 >> p
\end{equation}

It should be noted again that the bound $ n_0$ which appears in the
series over $ n$ is, in general, dependent upon the radial
variables, in addition to being dependent on $ p$. Consequently, in
(52) it is $ n_0' \neq n_0$ since the bound $ n_0$ which appears in
the singular part of the propagator in (44) depends on $ \rho_2$
whereas the bound $ n_0'$ relates exclusively to the value $ \rho_2
= \beta$. The physical significance of this difference will become
obvious in the next section.

Multiplying both sides of (52) by $ e^{-
i\frac{p'}{\beta}\tau_2}Y_{l'm'}^*(\theta_2, \phi_2)$ and
integrating yields

$$
\sum_{p = 0}^{\infty}\sum_{l = 0} ^{\infty}\sum_{m = -l}^{l}\sum_{n
= 1}^{\infty}A_{plmn}\frac{\beta^{p - 1}\Gamma(p +
1)}{\Big{[}i\sqrt{l^2 + l + 1}\Big{]}^p}J_p(2i\sqrt{l^2 + l +
1})\times
$$

$$
\int_0^{2\pi\beta}d\tau_2 e^{i\frac{p -
p'}{\beta}\tau_2}\int_0^{2\pi}d\phi_2\int_{-
1}^{1}dcos\theta_2Y_{lm}(\theta_2, \phi_2)Y_{l'm'}^*(\theta_2,
\phi_2) =
$$

$$
-
\frac{1}{\beta^{\frac{3}{2}}}\frac{1}{\sqrt{\rho_1}}\sum_{l=0}^{\infty}\sum_{m=-l}^lY_{lm}^{*}(\theta_1,
\phi_1)\int_0^{2\pi}d\phi_2\int_{-
1}^{1}dcos\theta_2Y_{lm}(\theta_2, \phi_2)Y_{l'm'}^*(\theta_2,
\phi_2)\times
$$

\begin{equation}
\sum_{p =
0}^{\infty}e^{-i\frac{p}{\beta}\tau_1}\sum_{n=n'_0[p]}^{\infty}\frac{cos[\frac{\pi}{4\beta}(4n
+ 2p + 3)(\beta - \rho_1)]}{\pi^2n^2 + 4(l^2 + l +
1)}\int_0^{2\pi\beta}d\tau_2 e^{i\frac{p - p'}{\beta}\tau_2}
\end{equation}
whereupon use of the orthonormality relation

\begin{equation}
\int_0^{2\pi}d\phi\int_{- 1}^{1}dcos\theta Y_{l'm'}^*(\theta ,
\phi)Y_{lm}(\theta , \phi) = \delta_{ll'}\delta_{mm'}
\end{equation}
as well as of

\begin{equation}
\int_0^Ldxe^{i\frac{2\pi}{L}(n - m)x} = L\delta_{mn}
\end{equation}
results in

$$
\sum_{n = 1}^{\infty}A_{plmn}\frac{\beta^{p - 1}\Gamma(p +
1)}{\Big{[}i\sqrt{l^2 + l + 1}\Big{]}^p}J_p(2i\sqrt{l^2 + l + 1}) =
$$

\begin{equation}
-
\frac{1}{\beta^{\frac{3}{2}}}\frac{1}{\sqrt{\rho_1}}e^{-i\frac{p}{\beta}\tau_1}
Y_{lm}^*(\theta_1, \phi_1)\sum_{n =
n_0'[p]}^{\infty}\frac{cos[\frac{\pi}{4\beta}(4n + 2p + 3)(\beta -
\rho_1)]}{\pi^2n^2 + 4(l^2 + l + 1)} ~~~;~~~ n_0' >> p
\end{equation}

In effect, the expansion coefficients in (51) are

\begin{equation}
 A_{plmn} =
 \left\{
\begin{array}{ll}
 0 ~~~;~~~ n < n_0' ~~~;~~~ n_0' >> p \\
 - \frac{1}{p!\beta^{p + \frac{1}{2}}}\frac{1}{\sqrt{\rho_1}}\frac{[i\sqrt{l^2 + l +
1}]^p}{J_p(2i\sqrt{l^2 + l + 1})}Y_{lm}^*(\theta_1,
\phi_1)e^{-i\frac{p}{\beta}\tau_1}\frac{cos[\frac{\pi}{4\beta}(4n + 2p + 3)(\beta -
\rho_1)]}{\pi^2n^2 + 4(l^2 + l + 1)} ~~~;~~~ n \geq n_0' \\
\end{array}
 \right.
\end{equation}

Replacing (57) in (51) results in

$$
D_{b}(x_2 - x_1) = - \frac{2}{
\beta^{\frac{3}{2}}}\frac{1}{\sqrt{\rho_1}}\times
$$

$$
\sum_{l = 0} ^{\infty}\sum_{m = -l}^{l}\sum_{p = 0}^{\infty}\sum_{n
= n_0'[p]}^{\infty}\frac{cos[\frac{\pi}{4\beta}(4n + 2p + 3)(\beta -
\rho_1)]}{\pi^2n^2 + 4(l^2 + l +
1)}\frac{J_p(\frac{2i}{\beta}\sqrt{l^2 + l +
1}\rho_2)}{J_p(2i\sqrt{l^2 + l + 1})}e^{i\frac{p}{\beta}(\tau_2 -
\tau_1)}Y_{lm}(\theta_2, \phi_2)Y_{lm}^*(\theta_1, \phi_1) ~~~;~~~
$$

\begin{equation}
n_0' >> p
\end{equation}


Again, in order to arrive at the expression which corresponds to the
actual infinite volume the series over $ n$ in (58) must be replaced
by an integral. Consequently, the consistent expression for the
boundary part $ D_b$ on the Euclidean black-hole geometry is

$$
D_{b}(x_2 - x_1) = - \frac{2}{
\beta^{\frac{3}{2}}}\frac{1}{\sqrt{\rho_1}}\times
$$

$$
\sum_{l = 0} ^{\infty}\sum_{m = -l}^{l}\sum_{p =
0}^{\infty}\int_{u_0'[p]}^{\infty}du\frac{cos[\frac{\pi}{4\beta}(4u
+ 2p + 3)(\beta - \rho_1)]}{\pi^2u^2 + 4(l^2 + l +
1)}\frac{J_p(\frac{2i}{\beta}\sqrt{l^2 + l +
1}\rho_2)}{J_p(2i\sqrt{l^2 + l + 1})}e^{i\frac{p}{\beta}(\tau_2 -
\tau_1)}Y_{lm}(\theta_2, \phi_2)Y_{lm}^*(\theta_1, \phi_1) ~~~;~~~
$$

\begin{equation}
u_0' >> p ~~~;~~~ \frac{\pi u}{\beta}\rho_{1} >> p
\end{equation}

In the context of (16) this is the expression for the boundary part
of the scalar propagator on the Euclidean Schwarzschild black-hole
geometry.

{\bf V. Scalar Propagation on the Black-Hole Geometry}\\

Within the range of radial values which are consistent with (16) the
thermal scalar propagator for a massless conformal scalar field on
the Euclidean Schwarzschild black-hole geometry is the sum total of
(47) and (59). As announced in (I.1), that is

$$
D(x_2 - x_1) =
$$

$$
\frac{2}{\beta}\frac{1}{\sqrt{\rho_1\rho_2}}\sum_{l=0}^{\infty}\sum_{m=-l}^{l}Y_{lm}(\theta_2,
\phi_2)Y_{lm}^*(\theta_1, \phi_1)\sum_{p =
0}^{\infty}e^{i\frac{p}{\beta}(\tau_2 -
\tau_1)}\int_{u_0[p]}^{\infty}du\frac{cos[\frac{\pi}{4\beta}(4u + 2p
+ 3)(\rho_2 - \rho_1)]}{\pi^2u^2 + 4(l^2 + l + 1)}
$$

$$
- \frac{2}{\beta^{\frac{3}{2}}}\frac{1}{\sqrt{\rho_1}}\times
$$

$$
\sum_{l = 0} ^{\infty}\sum_{m = -l}^{l}\sum_{p =
0}^{\infty}\int_{u_0'[p]}^{\infty}du\frac{cos[\frac{\pi}{4\beta}(4u
+ 2p + 3)(\beta - \rho_1)]}{\pi^2u^2 + 4(l^2 + l +
1)}\frac{J_p(\frac{2i}{\beta}\sqrt{l^2 + l +
1}\rho_2)}{J_p(2i\sqrt{l^2 + l + 1})}e^{i\frac{p}{\beta}(\tau_2 -
\tau_1)}Y_{lm}(\theta_2, \phi_2)Y_{lm}^*(\theta_1, \phi_1) ~~~;~~
$$

\begin{equation}
u_0 >> p ~~~~;~~~ u_0' >> p ~~~;~~~ \frac{\pi u}{\beta}\rho_{2,1}
>> p
\end{equation}

The significance of the range of values in (16) is that for $ \rho^2
<< \beta^2$ the solution to (17) approximates the solution to (15)
with an accuracy which increases with decreasing values of $ \rho$.
For that matter, the closer to the Schwarzschild event horizon the
propagation occurs the higher the accuracy with which the Green
function in (60) approximates the exact scalar propagator is. Such
considerations are qualitative. In order to make a quantitative
estimation of the accuracy of (60) it must be observed that the
operation of reducing (15) to (17) amounts essentially to treating $
\rho^2$ as a perturbation applied to the unknown solution to (15).
For that matter, the mathematical validity of such a reduction is
contingent upon the stability of that solution. However, (17) is a
linear differential equation whose solution (19) is bounded and
stable. The linear character of (15) ensures, for that matter, that
- at least in the range of values given by (16) and in the context
of the same boundary conditions of regularity at the origin and of a
vanishing amplitude at $ \rho = \beta$ - the solution to (15) is
also stable. Moreoever, the stability of the solution to (15) is
independently expected on the physical grounds that it corresponds
to the radial sector of the wave equation in the Schwarzschild
space-time. In effect, the reduction of (15) to (17) is
mathematically consistent.

In order to assess "how far" from the event horizon the radial
values can be considered before the Green function in (60) begins
to, substantially, detract from the expression of the exact
propagator it must be noticed that since, in the context of (16),
discarding $ \rho^2$ in favour of $ \beta^2$ does not affect the
stability of the solution to (15) values of $ \rho^2$ characterised
by two orders of magnitude below the value of $ \beta^2$ can be
safely considered to be consistent with (16). Consequently, the
range of validity for the Green function in (60) is

\begin{equation}
0 \leq \rho_i^2 \leq \frac{\beta^2}{100} ~~~~ ; ~~~~ i = 1, 2
\end{equation}
This estimate is physically reasonable. In fact, it will be
independently rederived in the context of the renormalisation of $
<\phi^2(x)>$ in section VI. That different approach will rigorously
establish the range stated in (61).

The estimate for the upper bound given in (61) for the radial
variable on the Euclidean sector of the Schwarzschild metric
translates to

\begin{equation}
2M \leq r \leq 2.0050M
\end{equation}
for the Schwarzschild radial coordinate in real time. This range
places an upper bound for $ r$ of the order of $ 10^{-3}r_S$ in
excess of $ r_S$, the radius of the Schwarzschild event horizon. For
a Schwarzschild black hole of one solar mass this amounts, in terms
of coordinate radial distance, to $ r = 3,006km$ - that is, to $ 6m$
"above" the event horizon. In order to assess the physical
significance of this range it must be recalled that the average
wavelength of the quanta emitted in the vicinity of the horizon is $
\sim M$. For that matter, in the Hartle-Hawking vacuum state, the
range in (62) explicitly refers to the segment of the static region
which is of central importance to particle creation. It is desirable
to know how large the range in (62) is with respect to the vacuum
activity itself. Such a determination requires a comparison with a
specific range of physical interest. The problem in this respect is
that since it is impossible to localise a quantum to within one
wavelength it also is meaningless to trace the origin of the emitted
particles to any particular region near the event horizon
\cite{Birrel}. Consequently, the range in (62) explored in the
Hartle-Hawking vacuum state does not determine whether the scalar
propagation in (60) is relevant to distances which extend far above
the event horizon in the exterior geometry. A physically valid way
to establish a comparison with a range of physical interest can be
obtained by invoking instead the situation of an observer who
accelerates to remain stationary at a finite $ r > 2M$. Unlike the
observer in free fall - who, for that matter, registers the
Hartle-Hawking vacuum state - the stated stationary observer will
register the emission of particles in the vicinity of the event
horizon. Since a black hole of one solar mass has a temperature of $
6\times 10^{-8}$K the stated observer registers the emission of
particles at proper distances which, for a black hole of one solar
mass, correspond to $ r \sim 10^{-6}$m above $ r = 2M = 3\times
10^{3}$m. This range coincides with that within which gravitational
back-reaction effects are substantial. The probability for particle
detection above this range rapidly decreases \cite{BryceDW}. Since -
for a Schwarzschild black hole of one solar mass - the upper bound
$r \sim 2.0050M$ in (62) corresponds to a coordinate radial distance
of $ 6$m above the event horizon that bound signifies a range of
validity for the propagator in (60) which corresponds to several
orders of magnitude above the range within which particles are
spontaneously created and back-reaction effects are pronounced. It
can be seen, for that matter, that there is an ample range of values
of $ r$ above $ r = 2M$ within which (61) and (62) signify
respectively a good approximation to the exact propagator.

The result expressed in (60) is substantially different from all
other approximate expressions which have been, hitherto, obtained
for the scalar propagator in various contexts. It signifies the only
approximation which specifically expresses the conformal scalar
propagator in terms of its space-time dependence in a specific
region of the Schwarzschild black-hole geometry.

In order to explore the implications of (60) to scalar propagation
in the vicinity of the event horizon it is also necessary to,
accurately, specify the range of possible values for the lower
bounds $ u_0$ and $ u_0'$. In fact, as a consequence of the
necessary condition

\begin{equation}
\frac{\pi u}{\beta}\rho_{1,2} >> p
\end{equation}
the lower bound $ u_0$ increases indefinitely as either $ \rho_2$ or
$ \rho_1$ tends to zero. The precise manner in which $ u_0$ depends
on $ \rho_2$ or $ \rho_1$ is central to the singularity structure of
the Green function in (60) and will be, accordingly, analyzed in
this section. Such considerations are also valid for $ p = 0$ since
the replacement of $ p$ by $ p + 1$ leaves (63) intact. Similar
considerations also apply to $ u_0^{'}$ as a function of $ \rho_1$.

The consequences of such an effect to propagation in the vicinity of
the event horizon can be revealed by first considering the integral
in the singular part of the propagator in (60). It is

\begin{equation}
\Big{|}\int_{u_0[p]}^{\infty}du\frac{cos[\frac{\pi}{4\beta}(4u + 2p
+ 3)(\rho_2 - \rho_1)]}{\pi^2u^2 + 4(l^2 + l + 1)}\Big{|} <
\frac{1}{\pi^2}\int_{u_0[p]}^{\infty}\frac{du}{u^2} =
\frac{1}{\pi^2}\frac{1}{u_0[p]}
\end{equation}
The relations in (64), (63) and (46) imply that

\begin{equation}
|D_{as}(x_2, x_1)| <<
\frac{1}{2\pi^2}\frac{1}{\beta^2}\frac{\sqrt{\rho_2}}{\sqrt{\rho_1}}\sum_{k
= 0}^{\infty}\sum_{p \neq 0}^{\infty}e^{i\frac{p}{\beta}(\tau_2 -
\tau_1)}\frac{2k + 1}{p}P_k(cos\gamma)
\end{equation}

It can be seen, for that matter, that since $ \sqrt{\rho_2}$
multiplies each term in the series the singular part $ D_{as}(x_2 -
x_1)$ of the propagator vanishes at $ \rho_2 \rightarrow 0$ on
condition that $ \rho_1$ be fixed at any value above zero. At once,
the boundary part $ D_b(x_2 - x_1)_{|\rho_2 = 0}$ can be seen to
remain at a non-vanishing value.

The same vanishing effect for the singular part can also be,
independently, obtained through use of the identity

\begin{equation}
\int_{u_0[p]}^{\infty}du\frac{cos[\frac{\pi u}{\beta}(\rho_2 -
\rho_1)]}{u^2} = \frac{1}{2}\pi^2\frac{\rho_2 - \rho_1}{\beta} +
\frac{1}{u_0}{}_1F_2[- \frac{1}{2} ~~~;~~~ \frac{1}{2},\frac{1}{2}
~~~;~~~ - \frac{\pi^2(\rho_2 - \rho_1)^2}{4\beta^2}u_0^2]
\end{equation}
with $ {}_1F_2$ being the generalised hypergeometric function which
admits the expansion

$$
{}_1F_2[- \frac{1}{2} ~~~;~~~ \frac{1}{2},\frac{1}{2} ~~~;~~~ -
\frac{\pi^2(\rho_2 - \rho_1)^2}{4\beta^2}u_0^2] = 1 + \frac{(-
\frac{1}{2})}{\frac{1}{2}\frac{1}{2}}[- \frac{\pi^2(\rho_2 -
\rho_1)^2}{1!\times (2\beta)^2}u_0^2] +
$$

\begin{equation}
\frac{(-\frac{1}{2})(1 - \frac{1}{2})}{\frac{1}{2}(1 +
\frac{1}{2})\frac{1}{2}(1 + \frac{1}{2})}[\frac{\pi^4(\rho_2 -
\rho_1)^4}{2!\times (2\beta)^4}u_0^4] + ...
\end{equation}
Again, at the limit $ \rho_2 \rightarrow 0$ (66) yields

\begin{equation}
lim_{u_0 \rightarrow
\infty}\int_{u_0[p]}^{\infty}du\frac{cos[\frac{\pi
u}{\beta}\rho_1]}{u^2} = 0
\end{equation}
which, through (64), signifies the same result as (65).

An immediate consequence of such an effect is that the Green
function in (60) is not defined at the limit $ \rho_2 \to 0$ when $
\rho_1 \neq 0$. The propagator vanishes in its entirety in such a
context eventhough its boundary part remains at a non-vanishing
value. In the absence of the singular part such a non-vanishing
expression is not, in any respect, associated with propagation. In
fact - as the preceding results reveal - the limit $ \rho_1 \to 0$
causes that expression to, also, vanish.

The vanishing effect expressed by

\begin{equation}
lim_{\rho_2 \rightarrow 0}D(x_2 - x_1) = 0 ~~~~ ; ~~~~ \rho_1 \neq 0
\end{equation}
is a consequence of the causal structure of the Schwarzschild
black-hole space-time. Specifically, at the semi-classical limit $
\hbar \to 0$ all observers in the static region, regardless of their
state of motion or choice of coordinates agree, that no particle
reaches the hole's event horizon within a finite advance of their
proper time. Equivalently, at the semi-classical limit all observers
in the static region agree that the frequency of a waveform tends to
zero in the vicinity of the hole's event horizon. Consequently, away
from the semi-classical limit the transition amplitude for quantum
propagation specified by one end-point of propagation being
arbitrarily close to the event horizon is also expected to vanish
for all observers in the static region. This situation is distinct
from that expressed by the single limit $ \rho \to 0 ~~ ; ~~ \rho_1
= \rho_2 \equiv \rho$ which is analysed below.

In order to explore the singularity structure of the propagator use
will be made of the fact that - as simple power counting reveals -
the singular part of the Green function in (60) contains a
logarithmic divergence in $ 1 + 1$ dimensions and a quadratic
divergence in four dimensions. Separating the space-time points in
the radial direction by setting $ x_1 = (\tau, \rho_1, \theta, \phi)
~~~~ ; ~~~~ x_2 = (\tau, \rho_1 + \epsilon_{\rho}, \theta, \phi)$,
and taking the limit $ \rho_2 \to \rho_1 \Rightarrow \epsilon_{\rho}
\to 0$ it is

\begin{equation}
D_{as}(x_2 \rightarrow x_1) =
\frac{1}{2\pi^3\beta}\frac{1}{\rho}\sum_{k = 0}^{\infty}(2k +
1)\Big{[}\sum_{p = 0}^{p_0
>> 1}\frac{1}{u_0} + \sum_{p = p_0}^{\infty}\frac{1}{u_0}\Big{]}
\end{equation}
where, in transfer space, use has also been made of (46) in the
ultra-violet domain $ u \rightarrow \infty$. The significance of
this domain is that, if the events $ x_2$ and $ x_1$ are arbitrarily
close to each other, the dominant contribution to the integral over
$ u$ comes from $ u >> l$ in each term in the series over $ l$.

The demand for a quadratic divergence with a logarithmic divergence
inherent in the radial-temporal sector of $ D_{as}$ imposes the
conditions

\begin{equation}
\left\{
\begin{array}{ll}
 \frac{1}{u_0} =
 \frac{1}{p}f^{}(\rho) + F_{kp}^{(1)}(\rho) ~~~~  ;
   ~~~~ ~~~ if  ~~~ p > p_0    \\
\frac{1}{u_0} = \frac{1}{v_p(\rho)}f^{}(\rho) + F_{kp}^{(2)}(\rho)
~~~~  ;
 ~~~ ~~~ if  ~~~ p < p_0   \\
v_p(\rho) \neq 0
\end{array}
 \right.
\end{equation}
for all values of $ p$ and $ \rho$ and on the understanding that
power counting in the ultra-violet domain precludes any dependence
of $ f(\rho)$ and $ v_p(\rho)$ on $ k$.

Replacing (71) in (70) yields

\begin{equation}
D_{as}(x_2 \rightarrow x_1) =
\frac{1}{2\pi^3\beta}\frac{1}{\rho}\sum_{k = 0}^{\infty}(2k +
1)\Big{[}\sum_{p = 0}^{p_0
>> 1}\frac{1}{v_p(\rho)} + \sum_{p =
p_0}^{\infty}\frac{1}{p}\Big{]}f(\rho) + F(\rho)
\end{equation}
and renders the quadratic divergence, which represents the
propagator's singularity at the coincidence space-time limit,
manifest.

The arbitrary term $ F(\rho)$ is finite. Its exclusive radial
dependence is expected on the grounds that the manifold is static
and spherically symmetric. As is also the case with the finite and
arbitrary sum over $ v^{-1}_p(\rho)$ the function $ F(\rho)$ emerges
as a necessary consequence of the operation of specifying the
divergence in (70). Its exact expression in the range stated in (61)
can only be independently determined by additional boundary
conditions in the corresponding physical context. Examples of such a
procedure will be given in what follows. It should be stressed,
however, that the finite character of $ F(\rho)$ is the exclusive
consequence of the behaviour of the scalar propagator at the
coincidence space-time limit and as such that character can not be
guarranted on the Schwarzschild black-hole's event horizon where the
regularity of a vacuum state - such as the Hartle-Hawking state $
|H>$ - in a freely falling frame is ensured only as a consequence of
the manner in which that state has been defined.

Since $ F_{kp}^{(1)}(\rho)$ and $ F_{kp}^{(2)}(\rho)$ are arbitrary
the quantity $ \frac{1}{u_0}$ can be redefined to be what it is in
(71) with $ F_{kp}^{(1)}(\rho) = F_{kp}^{(2)}(\rho) = 0$. In effect,
(63) reduces the - as of yet - remaining, unspecified functions in
(71) to being such that

\begin{equation}
\left\{
\begin{array}{ll}
 \frac{1}{u_0} =
 \frac{\pi}{\beta}
 \frac{1}{cp}\rho + F_{kp}^{(1)}(\rho) ~~~~  ;
   ~~~~ ~~~ if  ~~~ p > p_0    \\
\frac{1}{u_0} = \frac{\pi}{\beta} \frac{1}{v_p(\rho)}\rho +
F_{kp}^{(2)}(\rho) ~~~~  ;
 ~~~ ~~~ if  ~~~ p < p_0   \\
v_p(\rho) \approx cp ~~~~  ;  ~~~~ ~~~~ if ~~~ p >> 1  \\
\end{array}
 \right.
\end{equation}
with $ c > 1$ ; $ v_p(\rho) > 1 , \forall \rho$ and with $
[v_p(\rho)] \sim [m^0] ~~~ ; ~~~ [F_{kp}^{(1,2)}(\rho)] \sim
[m^{0}]$. The conditions in (73) explicitly reveal the information
which is inherent in (63) as to the singularity structure of the
propagator in (60). Specifically, as stated, the lower bound $ u_0$
increases indefinitely as either $ \rho_2$ or $ \rho_1$ tends to
zero. It can be seen from (70) and (71), for that matter, that if -
in the event that $ F_{kp}^{(1)}(\rho) = 0$ - the bound $ u_0$
behaves as $ \rho^{-k} ; k > 1 ; k \in Z$ or in any other respect
which causes $ \frac{1}{u_0}$ to be a non-linear function of $ \rho$
then the inverse $ \frac{1}{u_0}$ would cause each term in the
divergent infinite series over $ p$ in $ D_{as}(x_2 \rightarrow
x_1)$ to be multiplied by a function of $ \rho$. This situation is
physically inconsistent since a divergence - being the exclusive
consequence of the coincidence space-time limit $ x_2 \rightarrow
x_1$ - can only depend on local geometric quantities. This is, in
fact, an explicit feature of the representation of the Feynman
propagator for short space-time separations which the
Schwinger-DeWitt expansion yields. As a consequence, (63) imposes
the physical demand that - in the absence of the arbitrary function
$ F_{kp}^{(1)}(\rho)$ - $ k = 1$ and, hence, that $ \frac{1}{u_0}$
be a linear function of $ \rho$, on condition that $ p$ receive
values in the ultra-violet domain of transfer space. This demand is
manifestly expressed by the first relation in (73).

Away from the coincidence space-time limit $ x_2 \rightarrow x_1$
there is no reason that, in the absence of $ F_{kp}^{(2)}(\rho)$,
the expression of $ \frac{1}{u_0}$ be a linear function of $ \rho$.
For that matter, as a consequence of (63), the general expression of
that quantity for $ p < p_0$ is that in the second relation in (73).

The third relation, $ v_p(\rho) \approx cp$ for very large values of
$ p$, ensures that the second relation in (73) reduces to the first
as the transfer space variable $ p$ makes the transition from values
smaller than $ p_0 >>1$ to values bigger than $ p_0 >>1$.

The representation of the short distance behaviour of the exact
propagator by the Schwinger-DeWitt expansion suggests that the
underlying physical reason for the situation expressed in (73) is
that in the Schwarzschild geometry it is $ R_{\mu\nu}(x) = 0$ and $
R(x) = 0$.

In effect, the redefinition in (73) reduces (72) to

\begin{equation}
D_{as}(x_2 \rightarrow x_1) = \frac{1}{2\pi^2\beta^2}\sum_{k =
0}^{\infty}(2k + 1)\Big{[}\sum_{p = 0}^{p_0
>> 1}\frac{1}{v_p(\rho)} + \frac{1}{c}\sum_{p =
p_0}^{\infty}\frac{1}{p}\Big{]} + F(\rho)
\end{equation}

It should be remarked, in passing, that the quadratic divergence
which emerges by power counting at the coincidence space-time limit
is - at least as a leading divergence - a characteristic feature of
the scalar propagator on any manifold in four dimensions. This is
readily understood by the fact that the coincidence space-time limit
relates essentially to the Minkowski space tangent to the
pseudo-Riemannian manifold at the associated point. Consequently,
the background curvature does not have any effect on that
divergence.

On the event horizon (74) is

\begin{equation}
lim_{\rho \to 0}D(x_2 \rightarrow x_1) =
\frac{1}{2\pi^2\beta^2}\sum_{k = 0}^{\infty}(2k + 1)\Big{[}\sum_{p =
0}^{p_0
>> 1}\frac{1}{v_p(0)} + \frac{1}{c}\sum_{p =
p_0}^{\infty}\frac{1}{p}\Big{]} + F(0)
\end{equation}
since the boundary part vanishes at $ \rho_1 \to 0$.

In contrast to (69) the physical significance of (75) is that of the
dynamical behaviour of the quantum field at $ x_2 \to x_1$ evaluated
arbitrarily close to the event horizon in a freely falling frame.

In what follows these results will be exploited in order to
demonstrate the merit of the Green function in (60) in calculations
of local expressions.

{\bf VI. Evaluation of $ <\phi^2(x)>$ }\\

In real time the expectation value $ <\phi^2(x)>$ for the square of
the conformal scalar field in a specific vacuum state relates to the
Feynman propagator $ G_F(x_2 - x_1)$ corresponding to the same
vacuum state by

\begin{equation}
<\phi^2> = - i lim_{x_2 \to x_1} G_F(x_2 - x_1)
\end{equation}
In order to evaluate this expectation value in the Hartle-Hawking
vacuum state the divergent series in (74) must be accorded a pole
structure. This task has been accomplished in the Appendix. The
results therein allow for the evaluation of $ <\phi^2>$ on the
black-hole's event horizon. Replacing (A.10)

$$
\sum_{k = 0}^{\infty}(2k + 1) = \frac{1}{2}lim_{\epsilon \rightarrow
0^+}\frac{1}{\epsilon^{2l}}
$$
and (A.9)

$$
\sum_{k = 0}^{\infty}\sum_{p = p_0}^{\infty}\frac{2k + 1}{p} = -
\frac{1}{4}lim_{\epsilon \rightarrow 0^+}\frac{1}{\epsilon^{2l}}
~~~~ ; ~~~~ p_0 >> 1
$$
in (75) yields

\begin{equation}
lim_{\rho \to 0}D(x_2 \to x_1) = lim_{\epsilon \to 0^+}
\frac{1}{32\pi^2M^2}\frac{1}{c'(0)}\frac{1}{\epsilon^{2l}} + F(0)
\end{equation}
where

$$
\frac{1}{c'(\rho)} = \frac{1}{2}\sum_{p = 0}^{p_0 >>
1}\frac{1}{v_p(\rho)} - \frac{1}{4c} ~~~~ ; ~~~~ [c'(\rho)] \sim
[m^0]
$$

The information necessary for determining $ c'(0)$ and the free
parameter $ l$ in (77) is inherent in the nature of the
DeWitt-Schwinger expansion as an approximation to the exact
propagator for short separations. Specifically, in the context of
point splitting in real time $ t$ the space-time points are
separated along a geodesic in a non-null direction by an
infinitesimal distance \cite{Birrel}

$$
\sigma = \frac{1}{2}\sigma_{\mu}\sigma^{\mu} ~~~~ ; ~~~~\sigma^{\mu}
= 2\upsilon t^{\mu} ~~~~ ; ~~~~ t^{\mu}t_{\mu} = \pm 1
$$
For such separations the DeWitt-Schwinger expansion is a valid
approximation to the exact propagator - and, consequently, to the
Feynman propagator

\begin{equation}
G_F(x_2 - x_1) = iD(x_2 - x_1)
\end{equation}
in the context of (62). For that matter, the divergent - at the
coincidence space-time limit $ x_2 \to x_1$ - terms in the
DeWitt-Schwinger expansion coincide respectively, at the same limit,
with the divergent terms in $ G_F(x_2 - x_1)$. These considerations
constitute the basis for the renormalisation of the exact
propagator. The DeWitt-Schwinger approximation for short separations
is treated as a counterterm and subtracted from the unrenormalised
propagator before the coincidence space-time limit is taken. In
turn, the stated counterterm provides the basis for the one-loop
effective action corresponding to free propagation on the curved
manifold.

In the present context, where the mass of the black hole sets a
characteristic scale for the Schwarzschild metric, dimensional
analysis allows for

$$
\upsilon = M\epsilon
$$
and reveals that in (77) it is, necessarily, $ l = 1$ since the
singular term is already proportional to $ \upsilon^{-2}$. In fact,
for infinitesimal radial separations $ \epsilon_r = r_2 - 2M \equiv
M\epsilon^2$ from $ r_1 = 2M$ the demand that the divergence
contained in $ G_F(x_2 - x_1)$ coincide - at $ r_2 \to 2M$ - with
the divergent part \cite{Candelas}

\begin{equation}
\frac{1}{8\pi^2\sigma} = \frac{1}{32\pi^2M(r - 2M)} -
\frac{1}{192\pi^2M^2}
\end{equation}
stemming from the DeWitt-Schwinger expansion makes it obvious that
the limit $ \epsilon \rightarrow 0^+$ reduces to the limit $ \rho
\rightarrow 0 \Rightarrow r_2 \rightarrow 2M$ only when $ l = 1$ and
$ c'(0) = 1$.

As a result, (77) and (78) imply that

\begin{equation}
lim_{x_2 \to x_1}G_F(x_2 - x_1)_{|r_1 = 2M} = lim_{r_2 \to 2M}
\frac{i}{32\pi^2M(r_2 - 2M)} + ilim_{r_2 \to 2M}\tilde{F}(r_2)
\end{equation}
with $ \tilde{F}(r) = F[\rho(r)]$.

Replacing (80) in (76) yields

\begin{equation}
<H|\phi^2(x_2 ; r_2 = 2M)|H> = lim_{r_2 \to 2M}
\frac{1}{32\pi^2M(r_2 - 2M)} + lim_{r_2 \to 2M}\tilde{F}(r_2)
\end{equation}

It has already been stressed in section V that additional,
independent physical conditions are required on $ G_F(x_2 - x_1)$ in
order to determine the arbitrary term $ \tilde{F}(r_2)$. This task
can best be accomplished through the renormalisation of $
<H|\phi^2(x)|H>$ within the entire range given in (62).

The choice of temporal separations $ x_1 = (\tau, \rho, \theta,
\phi)$ and $ x_2 = (\tau + \epsilon_{\tau}, \rho, \theta, \phi)$ for
$ \rho \neq 0$ in the Euclidean sector of the metric followed by the
limit $ \epsilon_{\tau} \to 0$ taken on the expression in (60)
yields

$$
<H|\phi^2(x)|H> = \frac{1}{2\pi^2\beta^2}\sum_{k = 0}^{\infty}(2k +
1)\Big{[}\sum_{p = 0}^{p_0
>> 1}\frac{1}{v_p(\rho)} +
\frac{1}{c}\sum_{p = p_0}^{\infty}\frac{1}{p}\Big{]} + F(\rho)
$$

$$
- \frac{2}{\beta^{\frac{3}{2}}}\frac{1}{\sqrt{\rho}}\times
$$

$$
\sum_{l = 0} ^{\infty}\sum_{m = -l}^{l}\sum_{p =
0}^{\infty}\int_{u_0'[p]}^{\infty}du\frac{cos[\frac{\pi}{4\beta}(4u
+ 2p + 3)(\beta - \rho)]}{\pi^2u^2 + 4(l^2 + l +
1)}\frac{J_p(\frac{2i}{\beta}\sqrt{l^2 + l +
1}\rho)}{J_p(2i\sqrt{l^2 + l + 1})}|Y_{lm}(\theta, \phi)|^2 ~~~;~~
$$

\begin{equation}
u_0' >> p
\end{equation}
where (74), (76) and (78) have been used.

On the lines of the calculation which eventuated in (77) the
singular part reduces to

\begin{equation}
D_{as}(x_2 \to x_1) = lim_{\epsilon \to 0^+}
\frac{1}{32\pi^2M^2}\frac{1}{c'(\rho)}\frac{1}{\epsilon^{2k}} +
F(\rho)
\end{equation}

Repeating the point-splitting procedure for infinitesimal temporal
separations in real time $ \upsilon = \epsilon_{t} = M\epsilon$ - in
which procedure, pursuant to (1), it is $ t^{\mu}t_{\mu} = -1$ -
%
the demand that the divergence contained in $ G_F(x_2 - x_1)$ at $
\epsilon_t \to 0$ coincide with the divergence contained in the
DeWitt-Schwinger expansion to the relevant order \cite{Candelas}

\begin{equation}
\frac{1}{4\pi^2}\frac{1}{2\sigma} = \frac{1}{4\pi^2}\frac{1}{(1 -
\frac{2M}{r})\epsilon_t^2} + \frac{1}{4\pi^2}\frac{M^2}{12r^4(1 -
\frac{2M}{r})}
\end{equation}
at the same limit reveals that, in this context, it is $ k = 1$ and
$ \tilde{c}'(r) = c'[\rho(r)] = \frac{1}{8}(1 - \frac{2M}{r})$,
respectively. With these values (82) reduces to

$$
<H|\phi^2(x)|H> = \frac{1}{4\pi^2}lim_{\epsilon_{t} \to
0}\frac{1}{(1 - \frac{2M}{r})\epsilon_{t}^2} + \tilde{F}(r)
$$

$$
- \frac{1}{8M^2}\frac{1}{(1 - \frac{2M}{r})^{\frac{1}{4}}}\times
$$

$$
\sum_{l = 0} ^{\infty}\sum_{m = -l}^{l}\sum_{p =
0}^{\infty}\int_{u_0'[p]}^{\infty}du\frac{cos[\frac{\pi}{4}(4u + 2p
+ 3)(1 - \sqrt{1 - \frac{2M}{r}})]}{\pi^2u^2 + 4(l^2 + l +
1)}\frac{I_p[2\sqrt{(l^2 + l + 1)(1 -
\frac{2M}{r})}]}{I_p(2\sqrt{l^2 + l + 1})}|Y_{lm}(\theta, \phi)|^2
$$

\begin{equation}
u_0' >> p
\end{equation}
where use has been made of the relation

$$
I_p(x) = i^{-p}J_p(ix)
$$
between the modified Bessel function of the first kind $ I_p(x)$ and
the Bessel function $ J_p(ix)$.

The renormalisation of the divergent expression in (85) is
accomplished through the operation

\begin{equation}
<H|\phi^2(x)|H>_{ren} = lim_{\epsilon_{t} \to
0}\Big{[}<H|\phi^2(x)|H> - \frac{1}{8\pi^2\sigma}\Big{]}
\end{equation}
to the finite expression

$$
<H|\phi^2(x)|H>_{ren} = \frac{1}{12(8\pi M)^2} \frac{1 -
(\frac{2M}{r})^4}{1 - \frac{2M}{r}} - \frac{1}{12(8\pi
M)^2}\frac{1}{1 - \frac{2M}{r}} + \tilde{F}(r) - \frac{1}{32\pi
M^2}\frac{1}{(1 - \frac{2M}{r})^{\frac{1}{4}}}\times
$$

$$
\sum_{l = 0} ^{\infty}\sum_{p =
0}^{\infty}\int_{u_0'[p]}^{\infty}du\frac{cos[\frac{\pi}{4}(4u + 2p
+ 3)(1 - \sqrt{1 - \frac{2M}{r}})]}{\pi^2u^2 + 4(l^2 + l +
1)}\frac{I_p[2\sqrt{(l^2 + l + 1)(1 -
\frac{2M}{r})}]}{I_p(2\sqrt{l^2 + l + 1})}(2l + 1)
$$

\begin{equation}
u_0' >> p
\end{equation}
where following \cite{CandelHoward} the finite part of the
counterterm in (84) has been expressed as the difference of the
first two terms, the first of which remains finite throughout the
range of radial values stated in (62) whereas the other diverges at
$ r = 2M$. In (87) use has also been made of

$$
\sum_{m = -l}^lY_{lm}^*(\Omega')Y_{lm}(\Omega) = \frac{1}{4\pi}(2l +
1)P_l(cos\gamma)
$$
which is an immediate consequence of (46).

The underlying physical condition necessary to determine $
<H|\phi^2(x)|H>_{ren}$ is inherent in the Hartle-Hawking vacuum
state $ |H>$ itself as a consequence of whose definition the
renormalised stress-energy tensor $ <H|T_{\mu}^{\nu}(x)|H>_{ren}$ is
regular on both the past and the future horizon of the maximally
extended Kruskal manifold \cite{Birrel}, \cite{Candelas}. For that
matter, it immediately follows from (87) that

\begin{equation}
\tilde{F}(r) = \frac{1}{12(8\pi M)^2}\frac{1}{1 - \frac{2M}{r}} +
\tilde{F'}(r) ~~~ ; ~~~ \tilde{F'}(r) < \infty ~~~  \forall r \in
[2M, \infty)
\end{equation}
and, consequently, that

$$
<H|\phi^2(x)|H>_{ren} = \frac{1}{12(8\pi M)^2} \frac{1 -
(\frac{2M}{r})^4}{1 - \frac{2M}{r}} + \tilde{F'}(r) - \frac{1}{32\pi
M^2}\frac{1}{(1 - \frac{2M}{r})^{\frac{1}{4}}}\times
$$

$$
\sum_{l = 0} ^{\infty}\sum_{p =
0}^{\infty}\int_{u_0'[p]}^{\infty}du\frac{cos[\frac{\pi}{4}(4u + 2p
+ 3)(1 - \sqrt{1 - \frac{2M}{r}})]}{\pi^2u^2 + 4(l^2 + l +
1)}\frac{I_p[2\sqrt{(l^2 + l + 1)(1 -
\frac{2M}{r})}]}{I_p(2\sqrt{l^2 + l + 1})}(2l + 1)
$$

\begin{equation}
u_0' >> p ~~~; ~~~ 2M \leq r \leq 2.0050M
\end{equation}

The remaining $ \tilde{F'}(r)$ term is a regular function of $ r \in
[2M, \infty)$. For that matter, it is, itself, determined by the
fact that, at $ x_2 \to x_1$, (7) and (48) imply that the boundary
part of $ G_F(x_2, x_1)$ - which is featured in the third term of
(89) - necessarily corresponds to the addition of two functions of $
r$ one of which is $ \tilde{F'}(r)$ and the other is the finite sum
in (50). Since the boundary part, as a solution to (48), always
satisfies (51) it follows that in (89) it is

\begin{equation}
\tilde{F'}(r) = 0 ~~~; ~~~ 2M \leq r \leq 2.0050M
\end{equation}
Together (89) and (90) constitute the result announced in (I.2).

Similar considerations apply in the context of radial separations.
In the context of point splitting such separations, of course,
constitute the exclusive regulating approach when the coincidence
space-time limit is specified on the event horizon. Again, in view
of the counterterm in (79) the renormalisation of the divergent
expression in (81) is accomplished through the operation

\begin{equation}
<H|\phi^2(x ; r = 2M)|H>_{ren} = lim_{r_2 \to
2M}\Big{[}<H|\phi^2(x_2 ; r_2)|H> - \frac{1}{8\pi^2\sigma}\Big{]}
\end{equation}
to the finite expression

\begin{equation}
<H|\phi^2(x ; r = 2M)|H>_{ren} = \tilde{F}(2M) +
\frac{1}{192\pi^2M^2}
\end{equation}

The renormalised expression in (92) is regular on the event horizon
on condition that $ \tilde{F}(2M) < \infty$. In view of the fact
that

$$
lim_{r_2 \to 2M}\frac{1}{12(8\pi M)^2} \frac{1 - (\frac{2M}{r})^4}{1
- \frac{2M}{r}} = \frac{1}{192\pi^2M^2}
$$
of the fact that the boundary part of the propagator vanishes at $
r= 2M$ as well as of the fact that the expressions respectively
obtained for $ <H|\phi^2(x)|H>_{ren}$ through radial and temporal
separations are necessarily identical it follows from (89), (90) and
(92) that

\begin{equation}
\tilde{F}(2M) = 0
\end{equation}
and that consequently

\begin{equation}
<H|\phi^2(x ; r = 2M)|H>_{ren} = \frac{1}{192\pi^2M^2}
\end{equation}

The result in (94) is the renormalised value of $ <\phi^2(x)>$ on
the event horizon of the Schwarzschild black hole in the
Hartle-Hawking vacuum state. This result coincides with the
corresponding one cited in \cite{Candelas}.

In passing, it should be remarked that the physical condition of
regularity which eventuated in (88) is absent on the past event
horizon in the Unruh vacuum state and on both horizons in the
Boulware vacuum state. It is seen yet again, for that matter, that
the function $ F(\rho)$ which emerged in (72) at the coincidence
space-time limit necessarily depends on the choice of vacuum state.
In fact, the procedure followed above in the context of temporal and
radial separations is indicative of the manner in which the function
$ F(\rho)$ depends on both, the physical conditions set on a
hypersurface and the choice of regulating scheme. The extension of
the present results obtained in the context of the Hartle-Hawking
vacuum state to the corresponding contexts of the two other vacuum
states will be the object of future work.

Finally, following the coincidence of the result in (94) with that
in \cite{Candelas} consistency with established results requires
that the, analytically evaluated, renormalised expression in (89) be
compared with the results in \cite{CandelHoward} which have been
obtained through a combination of analytical and numerical
techniques and constitute an extension of the stated result in
\cite{Candelas} to the static region of the Schwarzschild black-hole
space-time. The result in \cite{CandelHoward} is of the form

\begin{equation}
<H|\phi^2(x)|H>_{ren} = \frac{1}{12(8\pi M)^2} \frac{1 -
(\frac{2M}{r})^4}{1 - \frac{2M}{r}} + \frac{\Delta(r)}{(8\pi M)^2}
\end{equation}
where the second term has been evaluated numerically.

In view of the fact that the first term in (95) identically
coincides with the first term in (89) and of the fact that, pursuant
to (90), the second term in (89) vanishes the second term in (95)
must be compared with the third term in (89). In order to accomplish
such a comparison it is necessary to change, in that term, the
independent variable $ r$ to the variable

\begin{equation}
\xi = \frac{r}{M} - 1
\end{equation}
used in \cite{CandelHoward}. The result is

$$
\frac{\tilde{\Delta}(r)}{(8\pi M)^2} = - \frac{2\pi}{(8\pi
M)^2}\Big{[}\frac{\xi + 1}{\xi - 1}\Big{]}^{\frac{1}{4}}\times
$$

\begin{equation}
\sum_{l = 0} ^{\infty}\sum_{p =
0}^{\infty}\int_{u_0'[p]}^{\infty}du\frac{cos[\frac{\pi}{4}(4u + 2p
+ 3)(1 - \sqrt{\frac{\xi - 1}{\xi + 1}})]}{\pi^2u^2 + 4(l^2 + l +
1)}\frac{I_p[2\sqrt{(l^2 + l + 1)\frac{\xi - 1}{\xi +
1}}]}{I_p(2\sqrt{l^2 + l + 1})}(2l + 1)
\end{equation}

In view of the fact that $ I_p(x)$ is an increasing function of $
x$, for all values of $ p$ and of the fact that $ 0 \leq \frac{\xi -
1}{\xi + 1} < 1$ ; $ lim_{\xi \to \infty}\frac{\xi - 1}{\xi + 1} =
1$ convergence of the expression in (97) can be rigorously
established by examining it in the context of the general asymptotic
expression \cite{Arfken}

\begin{equation}
I_{\nu}(z) \sim \frac{e^z}{\sqrt{2\pi z}} ~~~ ; ~~~ |z| >> \nu ~~~ ;
~~~ -\frac{\pi}{2} < Argz < \frac{\pi}{2}
\end{equation}

\begin{figure}[h]
\centering\epsfig{figure=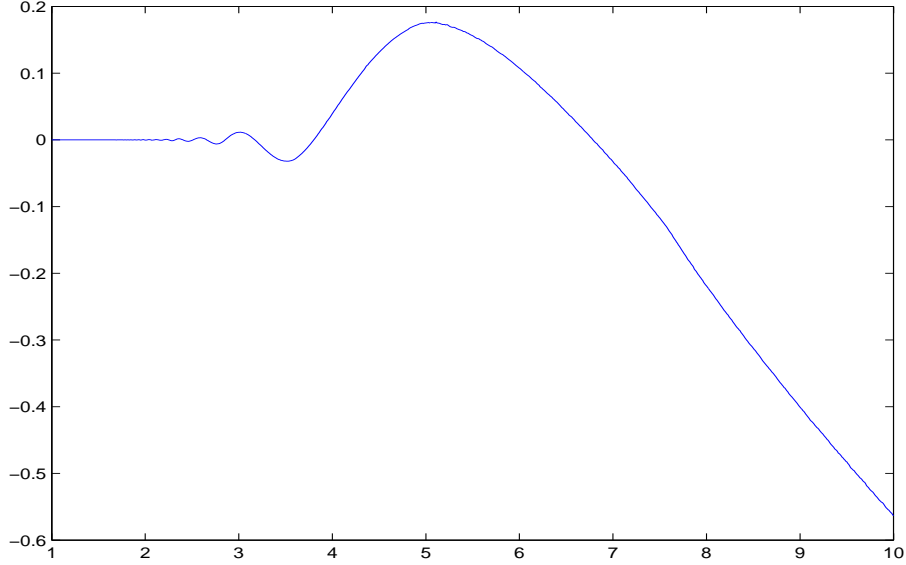, height = 76mm,width=120mm}
\caption{$ \tilde{\Delta}$ as a function of $ \xi$. The
approximation with $ \Delta$ emerges for values close to 1}
\end{figure}

The two graphs which will be analyzed in what follows have been
obtained through a program formulated in MATLAB. Although the graph
in Fig. 1 yields insufficient information as to the exact behaviour
of $ \tilde{\Delta}$ close to $ 1$ it manifestly features the stated
limit $ lim_{\xi \to 1}\tilde{\Delta}(\xi) = 0$. It can be seen that
there is an ample range of values of $ \xi$ for which the values of
$ \tilde{\Delta}$ are in conformity with the claim made in
\cite{CandelHoward} to the effect that for no values of $ \xi$ does
the second term in (95) exceed $ 1\%$ of the first. In fact,
inspection of the results cited in \cite{CandelHoward} for $
\Delta(\xi)$ reveals that the magnitude of the latter is of order
of, at most, $ 10^{-5}$ within the range $ 1 \leq \xi \sim 1.005$
which corresponds to that in (62).

\begin{figure}[h]
\centering\epsfig{figure=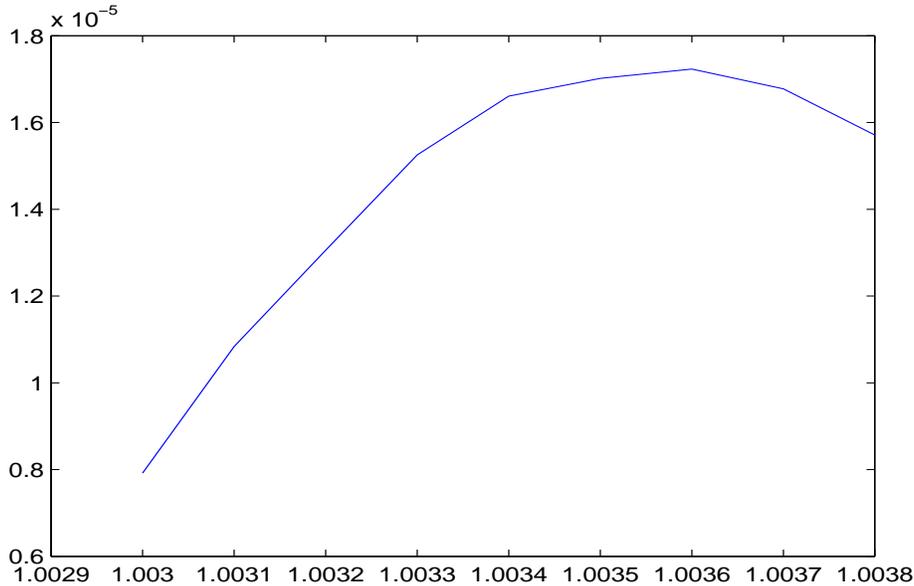, height = 85mm,width=141mm}
\caption{Behaviour of $ \tilde{\Delta}$ close to the Schwarzschild
event horizon. A consistent agreement with $ \Delta$ emerges for
values just under $ 1.005$}
\end{figure}

The graph in Fig. 2 manifestly expresses the behaviour of $
\tilde{\Delta}$ close to $ 1$. It can be seen that the range of
values for $ \tilde{\Delta}(\xi)$ corresponding to values of $ \xi$
between $ 1.003$ and $ \sim 1.004$ is in very good agreement with
the results indicated for $ \Delta(\xi)$ in \cite{CandelHoward} in
the same range of $ \xi$. In addition, the absence of any
oscillatory behaviour in that range suggests that $
\tilde{\Delta}(\xi)$ remains an increasing function also for values
of $ \xi$ between $ 1$ and $ 1.003$  and that, consequently, there
is very good agreement with the results in \cite{CandelHoward} for
values of $ \xi$ between $ 1$ and - at least - $ 1.004$. In effect,
Fig. 2 suggests that $ \tilde{\Delta}(\xi)$ increases from zero to a
value of order of, at most, $ 10^{-5}$ for values of $ \xi$ between
$ 1$ and $ \sim 1.005$, a range which corresponds to the physically
reasonable estimate in (62) for the validity of the approximation to
the scalar propagator. It is a limitation of the computer program
used herein that, for values of $ \xi$ between $ 1$ and $ 1.003$,
Fig. 2 reveals the stated information as to the behaviour of $
\tilde{\Delta}(\xi)$ only qualitatively. The exact behaviour of that
function in that range requires a more elaborate computer program.
Work in that direction is already in progress. However, the graphs
in Fig. 1 and Fig. 2 reveal a very good agreement - if not a
coincidence - between $ \tilde{\Delta}$ and $ \Delta$ at least
within the range of values specified in (62). Within the same range,
for that matter, the renormalised vacuum expectation value in (89)
is in very good agreement with the corresponding expression in (95).
This comparison renders the advantage which the Green function in
(60) has in the calculation of local expressions manifest. All the
information on the singularity structure of that Green function is
inherent in that function's singular part in (60). By properly
exploiting the universality of the quadratic divergence which a
scalar propagator has, at least as a leading divergence, in any
space-time that structure was rendered manifest in (74). At once,
the Green function in (60) is entirely analytic. In turn, through a
consistent renormalisation procedure $ <H|\phi^2(x)|H>_{ren}$ was
evaluated to the entirely analytic expression of the space-time
geometry in (89) with considerably less effort than that required
for the evaluation of the same physical quantity through the
combination of analytical and numerical approaches in
\cite{CandelHoward}.

In view of the fact that the range of the exterior geometry within
which (89) essentially coincides with (95) extends to the black
hole's event horizon the vacuum expectation value in (89) has the
superior feature of being entirely analytic in the most physically
important segment of the static region of the black-hole space-time
geometry. At once, the stated agreement between (89) and (95)
rigorously establishes the physically reasonable estimate made in
(61) for the same range in the Euclidean sector of the Schwarzschild
metric. That, is the range of physical relevance of the expression
obtained in (60) for the massless, conformal, scalar propagator on
the Schwarzschild black-hole geometry.

{\bf VII. Conclusions}\\

The approximation to the conformal scalar propagator associated with
the Hartle-Hawking vacuum state which has been developed herein
eventuated in an analytic expression which is, for that matter,
explicit in its dependence on the Schwarzschild black-hole
space-time. In effect, this approximation is sharply distinct from
all others. Although valid near the event horizon its range of
validity extends to several orders of magnitude above the range
within which quantum and backreaction effects are comparatively
pronounced. As a result of such an approximation certain aspects of
propagation related to the causal structure of the black-hole
space-time have been studied.

An essential advantage of the Feynman propagator which has been
developed herein is that its short-distance behaviour and
singularity structure are manifest. This propagator is, for that
matter, especially suited for an analytic evaluation of $
<T_{\mu\nu}(x)>$ and $ <\phi^2(x)>$. This aspect has been exploited
in order to reproduce established results for the renormalised value
of $ <\phi^2(x)>$ on the event horizon as well as in the segment of
the static region of the black-hole space-time which corresponds to
the range of validity of this approximation.

Although the evaluation of $ <T_{\mu\nu}(x)>$ in the static region
of the Schwarzschild black-hole and the manner in which this
compares to established results \cite{Fawcett}, \cite{Howard} is an
interesting enterprise an immediate priority is the extension of the
present results in the interior region of the Schwarzschild black
hole.

{\bf Acknowledgements}\\
I wish to acknowledge the inspiring effect of the ambient music of
sixteenth century composer John Taverner and of the electronic
ensemble "Tangerine Dream". I dedicate this work to the memory of my
father.

{\bf APPENDIX}\\

In this section the divergent infinite series in (74) will be
expressed as poles.

The logarithmically divergent series can be expressed as

$$
(A.1) \hspace{1in} \sum_{p = 1}^{\infty}\frac{1}{p} = lim_{x \to
\infty}\Big{[}(1 - \frac{1}{x}) + \frac{1}{2}(1 - \frac{1}{x})^2 +
\frac{1}{3}(1 - \frac{1}{x})^3 + ...\Big{]} \hspace{1in}
$$

In view of

$$
(A.2) \hspace{1.5in} -ln(1 - u) = u + \frac{u^2}{2} + \frac{u^3}{3}
+ ... ~~~~~; ~~~~~ 0 \leq u < 1 \hspace{2in}
$$
and of the fact that, prior to taking the limit $ x \to \infty$, the
condition for convergence in (A.2) is trivially satisfied (A.1)
yields

$$
(A.3) \hspace{1.5in} \sum_{p = 1}^{\infty}\frac{1}{p} = -lim_{x \to
\infty}ln[1 - (1 - \frac{1}{x})] = lim_{x \to \infty}lnx
\hspace{1in}
$$
Since the omission of a finite number of terms does not affect the
convergence properties of a series it, also, is

$$
(A.4) \hspace{2in} \sum_{p = p_0}^{\infty}\frac{1}{p} = lim_{x \to
\infty}lnx \hspace{2.5in}
$$

The quadratically divergent series in (74) requires a somewhat
different approach. Expressing it as

$$
(A.5) \hspace{1.5in} \sum_{k = 0}^{\infty}(2k + 1) = \Big{[}1 +
3x'^2 + 5x'^4 + ...\Big{]}_{|x'= 1} \hspace{2in}
$$
and setting

$$
g(x') = 1 + 3x'^2 + 5x'^4 + ...
$$
it is

$$
\int_0^{ < 1}dx'g(x') = \frac{x'}{1 - x'^2}
$$
so that

$$
g(x') = \frac{1 + x'^2}{(1 - x'^2)^2}
$$
as a result of which (A.5) becomes

$$
(A.6) \hspace{1in} \sum_{k = 0}^{\infty}(2k + 1) = lim_{x' \to
1^{-}}\frac{1 + x'^2}{(1 - x'^2)^2} = \frac{1}{2}lim_{x' \to
1^{-}}\frac{1}{(1 - x')^2} \hspace{1in}
$$
Further, setting

$$
x'= 1 - \frac{1}{x}
$$
results in

$$
(A.7) \hspace{2in} \sum_{k = 0}^{\infty}(2k + 1) = \frac{1}{2}lim_{x
\to \infty}x^2 \hspace{2in}
$$

Since (A.4) and (A.7) respectively imply that for an arbitrarily
large, but finite, value of $ x$ and for accordingly large values of
$ p_0$ and $ k_0$ it is

$$
\sum_{p = p_0}^{p'_0 >> p_0}\frac{1}{p} \sim lnx
$$
and

$$
\sum_{k = 0}^{k_0>>1}(2k + 1) \sim \frac{1}{2}x^2
$$
it follows that

$$
(A.8) \hspace{1.5in} \sum_{k = 0}^{\infty}\sum_{p =
p_0}^{\infty}\frac{2k + 1}{p} = \frac{1}{2}lim_{x \to \infty}x^2lnx
~~~~ ; ~~~~ p_0 >> 1 \hspace{2in}
$$
which through

$$
x = \frac{1}{\epsilon^l}
$$
finally becomes

$$
(A.9) \hspace{1.5in} \sum_{k = 0}^{\infty}\sum_{p =
p_0}^{\infty}\frac{2k + 1}{p} = - \frac{1}{4}lim_{\epsilon
\rightarrow 0^+}\frac{1}{\epsilon^{2l}} ~~~~ ; ~~~~ p_0 >> 1
\hspace{2in}
$$
where $ l$ is a parameter to be determined.

At once, (A.7) also implies

$$
(A.10) \hspace{1.5in} \sum_{k = 0}^{\infty}(2k + 1) =
\frac{1}{2}lim_{\epsilon \rightarrow 0^+}\frac{1}{\epsilon^{2l}}
\hspace{2.5in}
$$


\begin{thebibliography}{99}

\bibitem{Birrel} N. D. Birrell and P. C. W. Davies, {\bf Quantum Fields
in Curved Space}, Cambridge Monographs in Mathematical Physics,
(1984).
\bibitem{Christensen} S. M. Christensen, Phys. Rev. {\bf D14}, 2490,
(1976).
\bibitem{Anderson} P. R. Anderson, Phys. Rev. {\bf D41}, 1152,
(1990).
\bibitem{Candelas} P. Candelas, Phys. Rev. {\bf D21}, 2185, (1980).
\bibitem{AnHiSam} P. R. Anderson, W. A. Hiscock and D. A. Samuel,
Phys. Rev. {\bf D51}, 4337, (1995).
\bibitem{Page} D. N. Page,  Phys. Rev. {\bf D25},  1499, (1982).
\bibitem{SHawking} S. W. Hawking, Commun. Math. Phys. {\bf 80}, 421-442,
(1981).
\bibitem{GibbHawk} G. W. Gibbons and S. W. Hawking, Phys. Rev. {\bf
D15}, 2752, (1977).
\bibitem{Fawcett} M. S. Fawcett, Commun. Math. Phys. {\bf 89},
103-115, (1983)
\bibitem{BryceDW} A comprehensive treatment of Hawking radiation as
a form of the Unruh effect is contained in Bryce de Witt's article
{\bf Quantum gravity: the new synthesis} included in {\bf General
Relativity: An Einstein Centenary} edited by S. W. Hawking and W.
Israel, Cambridge University Press, (1979)
\bibitem{Arfken} G. Arfken, {\bf Mathematical Methods for
Physicists}, Academic Press, (1970)
\bibitem{CandelHoward} P. Candelas and K. W. Howard, Phys. Rev. {\bf
D29}, 1618, (1984)
\bibitem{Howard} K. W. Howard, Phys. Rev. {\bf D30}, 2532, (1984)
















\end{thebibliography}
\end{document}